\pdfoutput=1
\documentclass[10pt,journal,letterpaper,compsoc]{IEEEtran}
\usepackage{amsmath}
\usepackage{amssymb}
\usepackage{graphicx}
\usepackage{dcolumn}
\usepackage{bm}
\usepackage{url}

\newcommand{\rar}{\rightarrow}
\newcommand{\ra}{\rangle}
\newcommand{\la}{\langle}
\newcommand{\ttt}{\texttt}

\begin{document}
\title{The RDF Virtual Machine}

\author{Marko~A.~Rodriguez% <-this % stops a space
\IEEEcompsocitemizethanks{\IEEEcompsocthanksitem M.A. Rodriguez is with the Digital Library Research and Prototyping Team, the Center for Nonlinear Studies, and the Applied Mathematics and Plasma Physics Group of the Los Alamos National Laboratory, Los Alamos, New Mexico 87545.\protect\\%
E-mail: marko@markorodriguez.com Website: http://markorodriguez.com}}

\markboth{}%
{Rodriguez \MakeLowercase{\textit{et al.}}: The RDF Virtual Machine}

\IEEEcompsoctitleabstractindextext{%
\begin{abstract}
The Resource Description Framework (RDF) is a semantic network data model that is used to create machine-understandable descriptions of the world and is the basis of the Semantic Web. This article discusses the application of RDF to the representation of computer software and virtual computing machines. The Semantic Web is posited as not only a web of data, but also as a web of programs and processes.
\end{abstract}

\begin{IEEEkeywords}
Resource Description Framework, Virtual Machines, Distributed Computing, Semantic Web, Web Computing.
\end{IEEEkeywords}}

\maketitle

\IEEEdisplaynotcompsoctitleabstractindextext

\IEEEpeerreviewmaketitle

\section{Introduction}

At its core, the Semantic Web is a global graph data structure used to describe web resources in a machine-understandable way \cite{lee:semantic2001}. Unlike the World Wide Web, in which document resources are interconnected through a single type of relationship (i.e.~\ttt{href}, hyper-text links), on the Semantic Web, resources are related to one another through a heterogeneous set of relationships. The set of resources and relationship types are identified by Uniform Resource Identifiers (URI) \cite{uri:2001}. The Resource Description Framework (RDF) is a standard for graphing (i.e.~relating) URIs, literal values, and blank nodes (or anonymous nodes) \cite{rdfspec:manola2004}. If $U$ is the set of all URIs, $L$ is the set of all literal values, and $B$ is the set of all blank nodes, then an RDF triple (or link) is defined as $\la s, p, o \ra$, where $s \in (U \cup B)$, $p \in U$, and $o \in (U \cup L \cup B)$. The union of all triples constitutes the Semantic Web graph and can be generally defined as 
%%%
\begin{equation*}
G \subseteq \la (U \cup B) \times U \times (U \cup L \cup B) \ra.
\end{equation*}
%%%
At the level of RDF, the Semantic Web is simply a collection of triples. These triples form a data structure known as a directed edge labeled graph (or multi-relational network). However, in order to create a layer of abstraction to describe how resources should be interrelated and to reason about and infer non-explicit relationship between resources, ontological languages have been developed. The two most prevalent Semantic Web languages are the RDF Schema (RDFS) \cite{rdfs:brickley2004} and the Web Ontology Language (OWL) \cite{owlspec:mcguinness2004}. For a fine, practical review of these two languages see \cite{owl:lacy2005}.

The prevalent conception of the Semantic Web is that of a well-structured, massive-scale distributed data repository that can be utilized by applications for various purposes. However, the RDF data model is general enough to support not only the representation of data, but also the representation process. The purpose of this article is to discuss the general use of rule and process information in the Semantic Web and their explicit realization as RDF encoded software programs and RDF virtual machines (RVM). The remainder of this section will introduce 1.) the Linked Data initiative and its intention of creating a massive-scale distributed data structure and 2.) the RDF programming and virtual machine initiative and its intention of creating a massive-scale distributed process infrastructure. The latter initiative is a nascent movement which has the potential to greatly advance the utility of the Semantic Web and, as previously stated, forms the primary point of discussion for this article.

\subsection{Linked Data as a Distributed Data Structure}

The Linked Data initiative is concerned with exposing data within the URI address space much like the World Wide Web initiative is concerned with exposing documents and media in the URL address space \cite{berners:ldata2006}. Before discussing the Linked Data movement, it is important to understand how the Semantic Web serves not simply as a data repository, but more importantly as a single massive-scale distributed database. Moreover, it is important to discuss how the Semantic Web provides both a technological and cultural differentiation from the traditional notion of a database as posited by the relational database community. These factors set the Semantic Web up for being a revolutionary means by which data is globally managed and accessed.

Technologically, the Semantic Web is reminiscent of the relational database model, insofar as it is a data storage environment that provides well-structured data to external applications; though this data is not represented as a collection of interlinked tables, but instead as an edge labeled graph (more specifically, an RDF graph). While the table and graph data structures can be mapped into one another without loss of information, the utility of the graph structure has yielded the development of specialized graph databases known as triple stores \cite{lee:triple2004}.\footnote{It is important to note that many RDF servers still utilize an underlying relational database to manage data. Examples of such architectures include the D2R Server \cite{d2r:bizer2006}.} Moreover, the data exposed by the Semantic Web is within the URI address space and as such is agnostic to the addressing scheme of the underlying machine supporting its representation. In this way, the data on multiple physical machines are able to reference each other and thus, the Semantic Web serves as a single unified graph spanning serves worldwide.

Culturally, the Semantic Web maintains the open, globally accessible nature of the World Wide Web. In contrast, rarely are relational database schemas reused and/or openly distributed and rarely are relational database ports (e.g.~ODBC) made available for the public harvesting of information. The common paradigm in the relational database world is that data is accessed and manipulated by software with privileges to the data and only through that software is the information made available to other services, if at all. However, with respect to the Semantic Web, not only does the community encourage the distribution and reuse of ontologies\footnote{SchemaWeb available at \url{http://www.schemaweb.info/}.}, but it also provides open and accessible interfaces to the its data. Such interfaces are known as SPARQL endpoints \cite{sparql:prud2004} and HTTP-based linked RDF data \cite{berners:ldata2006}. The Semantic Web truly represents a new data management paradigm because of the way in which data is distributed and discovered: in an open, standards-based fashion. 

The Semantic Web's Linked Data community is focused on the systematic union of RDF datasets in order to allow
%%%
\begin{quote}
``[any man or machine] to start with one data source and then move through a potentially endless Web of data sources connected by RDF links. Just as the traditional document Web can be crawled by following hypertext links, the Web of Data can be crawled by following RDF links. Working on the crawled data, search engines can provide sophisticated query capabilities, similar to those provided by conventional relational databases. Because the query results themselves are structured data, not just links to HTML pages, they can be immediately processed, thus enabling a new class of applications based on the Web of Data." \cite{linkeddata:bizer2008}
\end{quote}
%%%
There is far-reaching potential for the Web of data that currently exists and will continue to grow to become. However, one of the limiting factors in the Linked Data approach is that while the community is providing a massive-scale distributed data structure, they are not providing a massive-scale distribute process infrastructure to compute on this Web of data \cite{rodriguez:distributed2008}. Without a distributed process infrastructure, Semantic Web applications are left with the typical server/client-download philosophy of the World Wide Web. For data intensive algorithms, this is an inefficient use of resources as it requires the movement of large amounts of data to the algorithm's executing machine(s). It is this design choice that has made the World Wide Web (i.e.~the web of HTML documents), at large, only accessible to those that have the processing power and space to download and index it.\footnote{A distributed process infrastructure is a feature of the Grid computing paradigm that provides a democratization of compute cycles \cite{grid:foster2004}. With respect to the Semantic Web, systems like GridVine provide a means to efficiently query and update an RDF graph that is overlaid across multiple physical machines \cite{gridvine:maur2007}.} For the keyword search space of the World Wide Web, this problem is perhaps best solved by the few large-scale, search engines in existence today. However, the Semantic Web, with its rich data model and nearly endless potential, is poised to require a new Web infrastructure to support its processing within and between its various Linked Data repositories. No single institution or organization will have the compute power, nor the man power, to execute and implement all the potentially useful algorithms that will make the Semantic Web stand out as the \textit{defacto} medium for representing data. In order to remedy this situation, a move towards a computing paradigm for the Semantic Web is necessary.

\subsection{RVM Computing as a Distributed Process Infrastructure}

The Semantic Web has the potential to not only act as a data storage environment, but also capture the more procedural aspects of computing, such as computer instructions and abstract virtual computing machines. In others words, given the flexibility of the RDF data model, it is possible to encode, in RDF, the rules by which RDF data is manipulated and thus, expose such information on the Semantic Web. Moreover, the URI address space is an infinite space that is only constrained by the size and number of physical machines that are supporting its representation. A flexible data model and an infinite address space make the Semantic Web an ideal medium for distributed, global computing. In this more computation-centric environment, instructions expressed in RDF are executed by RDF virtual machines (RVM). An RVM is any entity that processes RDF computing instructions, and in some instances, is represented in RDF as well. Thus, like other RDF data, computing instructions and RVMs are ``first-class" citizens on the Semantic Web. 

Many common computing models are made salient by the RVM paradigm, such as open (refer to Section \ref{sec:open-computing}), distributed (refer to Section \ref{sec:distributed-computing}), and reflective computing (refer to Section \ref{sec:reflective-computing}). RDF programming languages compile down to RDF and these RDF instructions can be accessed, annotated (i.e.~RDF related), and reasoned on like any other RDF data on the Semantic Web. Furthermore, unique situations emerge when RDF code is represented across different physical machines. Because all RDF instructions are in the same URI address space, there is nothing that prevents the software, much like the data, to by physically distributed. With an RDF virtual machine executing RDF instructions, it is possible for the virtual machine and the instructions to be relocated by simply downloading the RDF subgraph that represents that virtual machine to another physical machine. Thus, instead of migrating large amounts of data to a local environment for processing, the RDF virtual machine and the instructions that it is processing can be migrated to the remote environment. In this way, the process is moved to the data, not the data to the process. Finally, because RDF computing instructions and, in some cases, RVMs are represented in RDF, reflection is possible at the object, instruction, and machine-level. Thus, nearly the entire computing stack is exposed to reasoning and self-modifying processes.

The structure of this article is as follows. Section \ref{sec:rdfprogramming} discusses RDF-based programming languages in general and one language specifically: Neno \cite{rodriguez:gpsemnet2007}. Related work is also presented at the end of Section \ref{sec:rdfprogramming}. Section \ref{sec:triplecodervm} discusses the relationship between RDF computing instructions and RVMs and more specifically the Fhat and r-Fhat RVMs. Finally, Section \ref{sec:modelscomputing} discusses those aspects of computing on the Semantic Web -- open, distributed, and reflective computing -- that are conveniently exposed by the use of RDF-based programming languages, RVMs, RDF computing in general.

\section{RDF-Based Programming Languages}
\label{sec:rdfprogramming}

RDF is used to express facts about the world in a structured and machine understandable fashion.
%%%
\begin{quote}
``The basic intuition of model-theoretic semantics is that asserting a sentence makes a claim about the world: it is another way of saying that the world is, in fact, so arranged as to be an interpretation which makes the sentence true. In other words, an assertion amounts to stating a constraint on the possible ways the world might be." \cite{rdfsem:hayes2004}
\end{quote}
%%%
However, RDF is more generally useful and need not be constrained to asserting facts about the ``world". In this article, RDF is used to represent computational data structures such as software (i.e.~a sequence of instructions) and machine state (i.e.~operand stacks, program counters, etc.). Common data structures such as lists, trees, and graphs in general are conveniently represented in RDF, as are programs of Turing complete languages \cite{compute:turing1937}. This section focuses on one Turing complete RDF-based programming language called Neno \cite{rodriguez:gpsemnet2007}.\footnote{Neno/Fhat is currently available at \url{http://neno.lanl.gov/}.} Other RDF-based programming languages include the functional, stack-based language Ripple \cite{ripple:shinavier2007}\footnote{Ripple is currently available at \url{http://ripple.fortytwo.net/}.}, the object-oriented FABL \cite{fabl:bureau2001}\footnote{FABL is currently available at \url{http://fabl.net/}.}, Adenosine\footnote{Unfortunately, Adenosine has no formal publications nor a currently existing homepage. However, there are discussions of it on various Semantic Web mailing lists as well as a project homepage that is available through the Internet Archive. This article will briefly discuss its formalisms to provide a more complete picture of RDF-programming.}, and Adenine\footnote{Adenine is available at \url{http://www.ifcx.org/wiki/Adenine.html}.} \cite{adenine:quan2003} languages. These languages, along with RDF toolkits, RDF-to-object mappers, and Web-based rule languages will be discussed in the related work section.

\subsection{The Neno RDF Programming Language}

Neno is an imperative programming language that takes an object-oriented perspective on the resources of the Semantic Web \cite{rodriguez:gpsemnet2007}. In Neno, the human readable/writable language's grammar is similar to popular object-oriented programming languages such as Java and C++. However, as opposed to typical object-oriented languages, many of the constructs of the Neno language were designed to take advantage of the RDF data model and the standardized means by which RDF data in queried and modified (e.g.~SPARQL \cite{sparql:prud2004} and SPARQL/Update \cite{sparqlupdate:seaborne2007}, respectively). The motivation for many of the language constructs is to overcome the impedance mismatch between the typical object-oriented data model and the RDF data model.

\begin{figure}[h!]
	\begin{center}
		\includegraphics[width=0.485\textwidth]{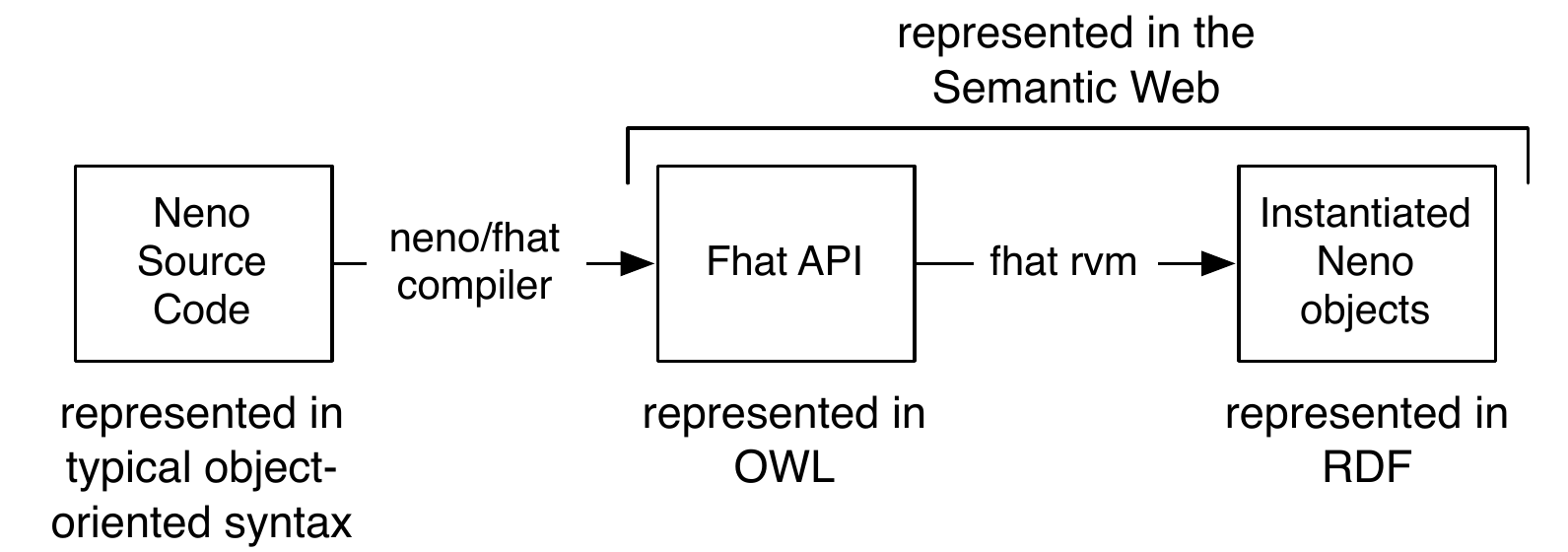}
	\caption{\label{fig:nenofhat-compilation}The various transformations from source code to computational object in Neno/Fhat.}
	\end{center}
\end{figure}

Neno source code is written by a human programmer and is compiled by the Neno/Fhat compiler. The compilation processes generates a Fhat API represented in OWL. A Fhat API denotes Neno classes, their respective methods, and each method's instructions. In this sense, a Fhat API is similar to the API of object-oriented languages (e.g.~the typical Java \ttt{jar} file). Classes in a Fhat API can be instantiated to active computational objects represented in RDF. These instantiated objects maintain low-level computing instructions (e.g.~\ttt{add}, \ttt{set}, \ttt{branch}, etc.) represented in RDF. These instructions denote computational primitives and specify the flow of execution within a method (refer to Section \ref{sec:fhat}). Figure \ref{fig:nenofhat-compilation} diagrams the the stages of processing required to go from human readable/writeable Neno source code to instantiated computational objects.
%%%

This sub-section will discuss the Neno programming language in particular and Section \ref{sec:fhat} will discuss the role of the Fhat RVM in the processing of compiled Neno source code.

\subsubsection{Neno Language Constructs}\label{sec:neno-constructs}

The following code example presents a simple Neno class and will be referred to throughout the remainder of this section.\footnote{For the sake of brevity, those operations that are typically found in other programing languages are not discussed in this section. For example, \ttt{for}-looping, \ttt{while}-looping, and \ttt{if/else} branching  have a similar syntax and behavior as found in other programming languages such Java and C. For an in-depth discussion of the Neno programming language constructs, please refer to \cite{rodriguez:gpsemnet2007}.}
%%%
%\begin{footnotesize}
%prefix owl: <http://www.w3.org/2002/07/owl>; 
%prefix xsd: <http://www.w3.org/2001/XMLSchema>; 
\begin{verbatim}
prefix lanl: <http://www.lanl.gov>; 
prefix foaf: <http://xmlns.com/foaf/0.1/>;

foaf:Agent lanl:Person {
  xsd:string foaf:name[1];
  lanl:Person foaf:knows[0..*];

  makeFriend(lanl:Person p) {
    this.foaf:knows =+ p;
  }

  makeEnemy(lanl:Person p) {
    this.foaf:knows =- p;
  }
  
  makeAllEnemies() {
    this.foaf:knows =/;
  }
  
  xsd:boolean isFriend(lanl:Person p) {
    return this.foaf:knows =? p;
  }
}
\end{verbatim}
%\end{footnotesize}
%%%
The \ttt{lanl:Person} class description is written in object-oriented syntax and is saved to a single text file denoted \ttt{Person.neno}. The class is composed of two fields and four methods and states that
%%%
\begin{itemize}
	\item \ttt{lanl:Person} is an \ttt{rdfs:subClassOf} of \ttt{foaf:Agent} (i.e.~extends),
	\item \ttt{foaf:name} is an \ttt{xsd:string} field that has a cardinality \ttt{owl:Restriction} of $1$,	
	\item \ttt{foaf:knows} is a \ttt{lanl:Person} field that does not have a cardinality \ttt{owl:Restriction},
	\item \ttt{makeFriend} is a void method that takes a single \ttt{lanl:Person} as an argument,
	\item \ttt{makeEnemy} is a void method that takes a single \ttt{lanl:Person} as an argument,
	\item \ttt{makeAllEnemies} is a void method that takes no arguments, and
	\item \ttt{isFriend} is an \ttt{xsd:boolean} method that takes a single \ttt{lanl:Person} as an argument.
\end{itemize}

Fields in Neno are assumed to be unordered sets because, for example, there may exist many \ttt{foaf:knows} relationship between two \ttt{lanl:Person} resources. As such, special set operators exist for interacting with Neno fields:
%%%
\begin{itemize}
	\item the \textit{set-plus} operator (\ttt{=+}): adds (i.e.~unions) a new value to a field.
	\item the \textit{set-minus} operator (\ttt{=-}): removes (i.e.~set minuses) an existing value from a field.
	\item the \textit{set-clear} operator (\ttt{=/}): removes all existing values from a field.
	\item the \textit{set-query} operator (\ttt{=?}): returns a boolean specifying whether the provided value currently exists in the field.
	\item the \textit{set} operator (\ttt{=}): removes all existing values and adds the provided values to the field.
\end{itemize}

Typical object-oriented ``dot" notation is used to reference the fields and methods of an object. For example, with respect to object fields, the statement
%%%
\begin{verbatim}
 lanl:marko.foaf:knows.foaf:name;
\end{verbatim}
%%%
returns the \ttt{xsd:string} \ttt{foaf:name}s of all resources that \ttt{lanl:marko} knows. That is, it first resolves all the \ttt{lanl:Person} instances that \ttt{lanl:marko} knows and then resolves all the \ttt{xsd:string} \ttt{foaf:name} values of those \ttt{lanl:Person}s. Neno also supports a non-standard ``dot dot" notation that is used for \textit{inverse referencing}. For example, the statement
%%%
\begin{verbatim}
 lanl:marko..foaf:knows.foaf:name;
\end{verbatim}
%%%
returns the names of all the resources that know \ttt{lanl:marko}. That is, it determines all the \ttt{lanl:Person}s that \ttt{foaf:know} \ttt{lanl:marko} and then returns the \ttt{xsd:string} of their \ttt{foaf:name}. The difference between ``dot" notation and ``dot dot" notation can be illustrated with two SPARQL queries. The first ``dot" statement translates to the query 
%%%
\begin{verbatim}
SELECT ?y 
  WHERE {
    <lanl:marko> <foaf:knows> ?x .
    ?x <foaf:name> ?y }
\end{verbatim}
%%%
and the second ``dot dot" statement translates to
%%%
\begin{verbatim}
SELECT ?y 
  WHERE {
    ?x <foaf:knows> <lanl:marko> .
    ?x <foaf:name> ?y }.
\end{verbatim}
%%%
The ``dot dot" notation takes advantage of the network structure of the underlying RDF data model and the ability to traverse that graph in any direction using query languages such as SPARQL. This is related to the ``ancestor" query mechanisms found in XPath \cite{xpath:clark1999} and used in semi-structured data environments such as Lorel \cite{lorel:abiteboul1997}.

Like typical object-oriented programming languages, ``dot" notation can be used to invoke an object's method. For example, suppose the \ttt{makeEnemy} method declaration for \ttt{lanl:Person}. The purpose of this method is to remove a \ttt{foaf:knows} relationships between the executing object (i.e.~\ttt{this}) and the provided \ttt{lanl:Person} parameter \ttt{p}.  Thus,
 %%%
 \begin{verbatim}
   lanl:marko.makeEnemy(lanl:dr_wh);
 \end{verbatim}
%%%
executes the following SPARQL/Update command:
%%%
\begin{verbatim}
DELETE {
  <lanl:marko> <foaf:knows> <lanl:dr_wh> }.
\end{verbatim}
%%%
In Neno, the ``dot dot" notation can also be applied to methods, and in such cases, it is called \textit{inverse method invocation}. Inverse method invocation can be used to remove all the \ttt{foaf:knows} relations between those \ttt{lanl:Person}s that \ttt{lanl:marko} \ttt{foaf:knows} and \ttt{lanl:dr\_wh}. In other words, all of Marko's friends can be instructed to make enemies with Dr. Wh. In Neno syntax, this is represented as
%%%
\begin{small}
\begin{verbatim}
 lanl:marko..foaf:knows.makeEnemy(lanl:dr_wh);
\end{verbatim}
\end{small}

While Neno has many similarities to typical object-oriented programming languages such as Java and C++, perhaps the most interesting aspect of Neno's programming constructs is the way in which it takes advantage of the underlying RDF representation of its instantiated objects. For a more in-depth review of the Neno programming languages which includes discussion of looping, branching, as well as object construction and destruction, please refer to \cite{rodriguez:gpsemnet2007}. Finally, note that Section \ref{sec:fhat} will discuss compiled Neno code and its representation in RDF.

\subsection{Related Work}

The ideas behind Neno come from a longline of Web-based process models. This subsection will provide a review of related work in this area with particular focus on other RDF programming languages, RDF toolkits, RDF-to-object mappers, and finally, other popular process description mechanisms for the Web.

\subsubsection{Other RDF Programming Languages}

Other known RDF programming languages include Ripple \cite{ripple:shinavier2007}, FABL \cite{fabl:bureau2001}, Adenosine, and Adenine \cite{adenine:quan2003}. These languages have a similar philosophy to Neno in that that they are motivated by the desire to encode both data and process information within RDF and thus, take unique advantage of the Semantic Web infrastructure. The general theme behind all of these languages is to turn the Web into a distributed computing environment.

Ripple is a declarative programming language aimed at Semantic Web mashups and scripting applications. In Ripple, human-readable programs expressed in a Notation3-like  \cite{n3:lee1998} serialization language are translated to and from RDF computing instructions in the form of linked RDF lists. The RDF lists which make up a Ripple script are intended to reside in the Semantic Web itself and thus, are at the same level of abstraction as the data they operate upon.

\begin{figure}[h!]
	\begin{center}
		\includegraphics[width=0.5\textwidth]{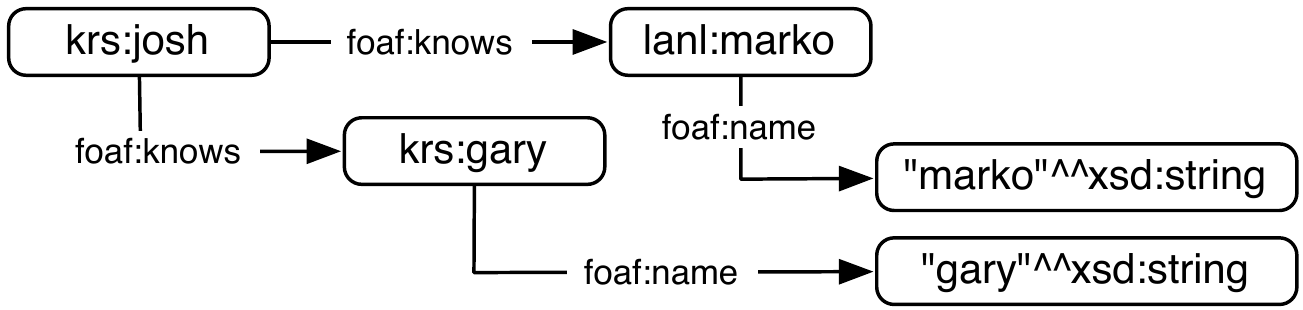}
	\caption{\label{fig:knows-network}An example RDF graph.}
	\end{center}
\end{figure}

Ripple is particularly useful for path-based traversals. For example, the Ripple query 
%%%
\begin{verbatim}
krs:josh foaf:knows! foaf:name!
\end{verbatim}
%%%
yields the name of all of the individuals that \ttt{krs:josh} knows. For example, consider the RDF graph illustrated in Figure \ref{fig:knows-network}. The above Ripple query has the effect of pushing both \ttt{"marko"$^\wedge$$^\wedge$xsd:string} and \ttt{"gary"$^\wedge$$^\wedge$xsd:string} onto the Ripple RVM stack.\footnote{The term RVM was introduced in \cite{rodriguez:gpsemnet2007} and refers to a virtual machine that processes RDF instructions. However, unlike the languages presented here, the Fhat RVM of \cite{rodriguez:gpsemnet2007} was also encoded in RDF. For the purpose of this article, both RDF and non-RDF represented virtual machines that process RDF instructions are called RVMs.} Once \ttt{"marko"$^\wedge$$^\wedge$xsd:string} and \ttt{"gary"$^\wedge$$^\wedge$xsd:string}  are on the stack, other operations can be performed on that data.
%%%

\begin{figure}[h!]
	\begin{center}
		\includegraphics[width=0.425\textwidth]{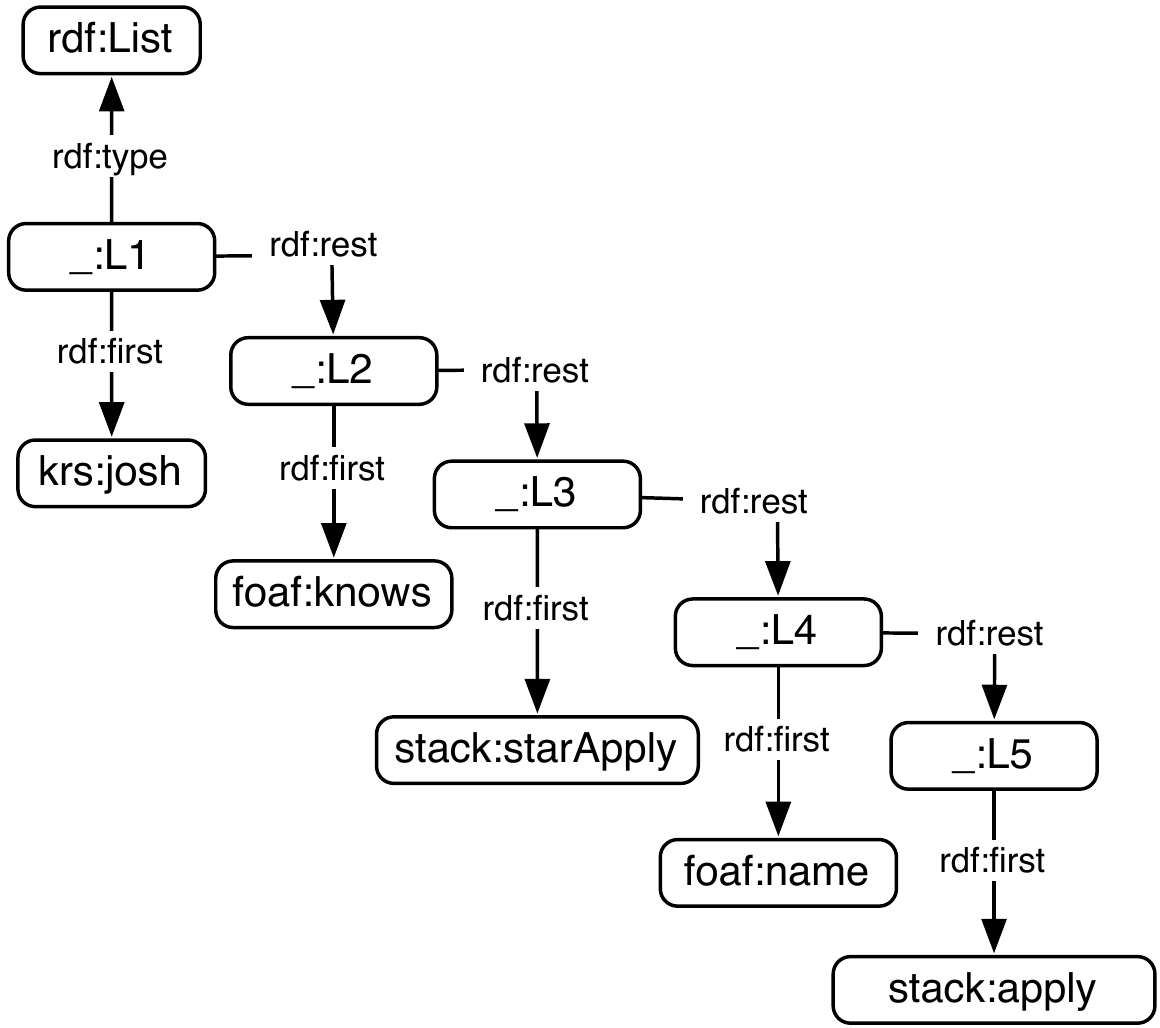}
	\caption{\label{fig:ripple-graph}An RDF representation of a compiled Ripple program.}
	\end{center}
\end{figure}

An interesting aspect of the Ripple language is its ability to ``walk" an RDF graph in a recursive fashion. For example, the following Ripple query
%%%
\begin{verbatim}
krs:josh foaf:knows* foaf:name!
\end{verbatim}
%%%
yields Josh's name, the names of those known by Josh, and so on, recursively. Figure \ref{fig:ripple-graph} diagrams the explicit RDF representation of this Ripple program, where \ttt{\_:L1}, \ttt{\_:L2}, \ttt{\_:L3}, \ttt{\_:L4}, and \ttt{\_:L5} are blank nodes of \ttt{rdf:type} \ttt{rdf:List}. 
%%%

FABL is an object-oriented language that has some similarities in syntax to JavaScript, compiles down to RDF computing instructions, and is executed by the FABL RVM. Like Neno, the native objects of FABL are RDF resources, however, the classes of FABL are DAML+OIL classes \cite{damloil:mcguinness2002}. An example of FABL code is:
%%%
\begin{verbatim}
allocate('foaf:knows',Property);

class("foaf:Person");
  restrict foaf:knows 
    {allValuesFrom foaf:Person}
endClass();

boolean function isFriend(foaf:Person p,
      foaf:Person q) {
  return contains(p..foaf:knows,q);
}
\end{verbatim}
%%%
The ``dot dot" notation in the \ttt{isFriend} method iterates over all the \ttt{foaf:knows} properties as properties are treated as sequences of values.

Adenosine is described by its originator as
%%%
\begin{quote}
[...] a language designed both to work on, and be distributed over, the Semantic Web. The language exploits the expressiveness of RDF whilst adopting a clean syntax based on a combination of Notation3 and ECMAScript.\footnote{Adenosine was previously available at \url{http://www.netalleynetworks.com/community/jgeldart/research/adenosine/}.}
\end{quote}
%%%
An example of the Adenosine syntax is:
%%%
\begin{verbatim}
lanl:isFriend a std:Method;
  std:onClass lanl:Person;
  std:function lanl:isFriendFunction.

@function lanl:isFriendFunction(p1, p2) {
  return p2 in p1.foaf:knows;
}
\end{verbatim}
%%%
Adenosine RDF code is executed by an RVM called Callaghan.

Finally, the \ttt{isFriend} method is demonstrated in Adenine. Note that both Adenine and Adenosine have a similar syntax, and in fact, Adenosine was developed after Adenine in order to provide (as decided by the designer of Adenosine) a better syntax. Hence the similar names that the two languages have.
%%%
\begin{verbatim}
add { lanl:Person
  rdfs:subClassOf foaf:Person ;
}

add { foaf:knows
  rdf:type rdf:Property ;
  rdfs:domain foaf:Person ;
  rdfs:range foaf:Person ;
}

method lanl:isFriend p1 p2
    return (contains p1 foaf:knows p2)
\end{verbatim}

\subsubsection{RDF Toolkits and RDF-to-Object Mappers}

The previously presented RDF programming languages serve a different purpose than RDF toolkits such as Jena \cite{jena:mcbride2002}\footnote{Jena is currently available at \url{http://jena.sourceforge.net/}.}, Sesame \cite{sesame:dutchy2002}\footnote{OpenRDF is currently available at \url{http://www.openrdf.org/}.}, Redland \cite{redland:beckett2001}\footnote{Redland is currently available \url{http://librdf.org/}.}, RDFStore\footnote{RDFStore currently available at \url{http://rdfstore.sourceforge.net/}.}, RDFLib\footnote{RDFLib is currently available at \url{http://rdflib.net/}.}, and Pyrple\footnote{Pyrple is currently available at \url{http://infomesh.net/pyrple/}.}. The purpose of these toolkits is to provide a mechanism by which RDF data can be accessed and manipulated through the constructs of a specific non-RDF-based programming language such as Java, C, PHP, Perl, Python, and/or Ruby. The difference between these languages and the RDF-based programming languages presented previous are that RDF programming languages are designed specifically to work with RDF and as such, can provide
%%%
\begin{itemize}
	\item type-checking at the RDF level, 
	\item data and process encapsulation,
	\item language operators to deal specifically with the RDF data model,
	\item no impedance mismatch between RDF and the manipulating language, and of specific differentiation, 
	\item can provide a representation of the procedural information within RDF.
\end{itemize}
%%%

RDF-to-object data mappers are related to RDF toolkits in that they aid a developer in utilizing RDF data in a programming language environment. Example RDF-to-object data mappers include Schemagen\footnote{Schemagen is currently available at \url{http://jena.sourceforge.net/}.}, Elmo\footnote{Elmo is currently available at \url{http://www.openrdf.org/}.}, Frege \cite{geldart:rdfrevolution2005}, and ActiveRDF \cite{activerdf:oren2008}. The purpose of an RDF-to-object data mapper is to alleviate the issues surrounding the impedance mismatch between the RDF data model and typical object-oriented data models. This is accomplished by 1.) automatically generating class definition in the non-RDF language that can interact with an RDF representation and 2.) automatically populate these objects using RDF data. With RDF-to-object mapping, what is preserved in the Semantic Web is the description of the data contained in an object (i.e.~object fields), not an explicit representation of the object's process information (i.e.~object methods). By explicitly encoding method data, the Semantic Web contains all the information required to retrieve and execute the behaviors of the object. In this way, with RDF programming languages, they do not require a separate, non-RDF programming environment to function. Moreover, the difficulties associated with RDF-to-object mappings \cite{owljava:kalyanpur2004,activerdf:oren2008} are not present in RDF-based programming languages. Some of the distinctions between the semantics of oriented-oriented programming languages and the semantics of RDF are made salient when understanding the distinction between frame-based languages and ontological languages such as OWL \cite{frameowl:wang2006}.

\subsubsection{Other Web-Based Process Descriptions}

Rule-based markup languages, designed for the World Wide Web and the Semantic Web, are a related area of research. Similar to RDF programming languages, the intent of this research is to formalize computing instructions within the data repository itself, whether that data repository be the World Wide Web and/or the Semantic Web. The standardization group focused on the W3C RuleML\footnote{RuleML is currently available at \url{http://www.ruleml.org/}.} initiative have developed XML languages for encoding process information as well as translators for representing this information in RDF \cite{rdfruleml:boley2005}. Particular variants of RuleML include a first order logic language (FOL-ML) \cite{folruleml:boley2005} and an object-oriented language (OO-ML) \cite{ooruleml:boley2003}. To provide an example of FOL-ML, the following logic statement 
%%%
\begin{equation*}
	\forall x.[\text{knows}(\text{marko},x) \implies \text{knows}(\text{dr\_wh}, x)].
\end{equation*}
%%%
states that for all the people that Marko knows, Dr. Wh also knows those people. This statement can be represented In FOL-ML as
%%%
\begin{verbatim}
<Forall>
  <Var>x</Var>
  <Implies>
    <Atom>
      <oid> 
        <Ind uri="lanl:marko"/> 
      </oid>
      <slot> 
        <Ind uri="foaf:knows"> 
        <Var>x</Var> 
      </slot>
    </Atom>
    <Atom>
      <oid> 
        <Ind uri="lanl:dr_wh"/> 
      </oid>
      <slot> 
        <Ind uri="foaf:knows"> 
        <Var>x</Var> 
      </slot>
    </Atom>
  </Implies>
</Forall>
\end{verbatim}

The various RuleML reasoners serve as  the virtual machines that execute RuleML documents. For example, jDrew and its object oriented variant OO jDrew \cite{ball:oojdrew2005} are deductive reasoning engines for RuleML.\footnote{jDrew is currently available at \url{http://www.jdrew.org/}.} With respect to rule-based systems designed specifically for the Semantic Web, there currently exists the Semantic Web Rule Language (SWRL) \cite{swrl:horrocks2004} which found its roots in RIF (Rule Interchange Format) \cite{rif:boley2008}. The Pellet reasoner currently supports SWRL rules \cite{pellet:2004}.\footnote{Pellet is currently available at \url{http://pellet.owldl.com/}.} Moreover, the Pellet reasoner along with other description logic reasoners can execute the reasoning rules of the OWL language. In this respect, the OWL language is a process description, albeit, it is not Turing complete. Also, there exist the Euler proof mechanism\footnote{Euler is currently available at \url{http://www.agfa.com/w3c/euler}.} for reasoning and Cwm\footnote{Cwm is currently available at \url{http://www.w3.org/2000/10/swap/doc/cwm}.} for general-purpose data processing on the Semantic Web. Finally, another area where process information is specifically encoded in the Semantic Web is the OWL-S service description framework \cite{owls:martin2004}. 

\section{RDF Instructions and the RVM}\label{sec:triplecodervm}

As the Semantic Web is simply a data structure and does not, in and of itself, have the ability to compute, external machines are required to manipulate its state by adding and removing triples. Even when state transition rules are explicitly encoded in the Semantic Web as machine instructions, there must still exist a computing machine that is able to process those instructions.

\subsection{An Introduction to Virtual Machines}

A virtual machine is a computing machine represented in software as opposed to a hardware (e.g.~logic gates) \cite{vm:craig2005}. There are many machines that fit this description and range in complexity from low-level VHDL machines \cite{vhdl:coelho1988} to the high-level interpreters of scripting languages such as Perl, JavaScript, and Python. Perhaps the most popular virtual machine is the Java virtual machine (JVM) of the Java programming environment \cite{jvm:lindholm1999}. There are two primary components to the Java environment: the Java compiler (i.e.~\ttt{javac}) and the JVM (i.e~\ttt{java}). The Java compiler translates human readable/writeable Java source code into Java byte-code (e.g.~\ttt{javac Person.java} $\rar$ \ttt{Person.class}). Java byte-code is executed by the JVM (e.g.~\ttt{java Person}). Each byte-code instruction alters the state of the JVM, whereby new variables are declared, changed, and ultimately carry out a user-defined computation. 

\subsection{An Introduction to RVMs}

An RVM is any virtual machine that interprets RDF computing instructions. Like typical virtual machines, RVMs can vary in the degree of detail that they formally represent. In the Ripple environment, the Ripple RVM's state and process rules are implemented in the Java language \cite{ripple:shinavier2007}. Thus, the Ripple RVM runs on the JVM.\footnote{It is possible, given that Ripple is Turing complete, to build a completely RDF-based virtual machine in Ripple and thus, encode both Ripple programs and the Ripple RVM in the Semantic Web as RDF computing instructions. This is also possible with the other RDF programming languages as they are all Turing complete.} In the Neno/Fhat environment, the Fhat RVM represents its state in RDF and the process by which that state is altered in Lisp \cite{rodriguez:gpsemnet2007}. Thus, in Neno/Fhat, not only are the RDF computing instructions encoded in the Semantic Web, but so is the state of the RVM (i.e.~its stacks, frames, program counter, etc.). The purpose of encoding an RVM state in RDF is to migrate RVMs between physical machines (refer to Section \ref{sec:distributed-computing}). However, note that there will always be a level of indirection in which computation is moved out of the Semantic Web to the physical hardware which supports it. In the end, it is the physical hardware that changes the state of the Semantic Web. Moreover, it is the laws of physics that drive the evolution of a hardware processor. Thus, in order to compute, every level of process abstraction must be grounded in (or founded on) some physical process.

\begin{figure}[h!]
	\begin{center}
		\includegraphics[width=0.375\textwidth]{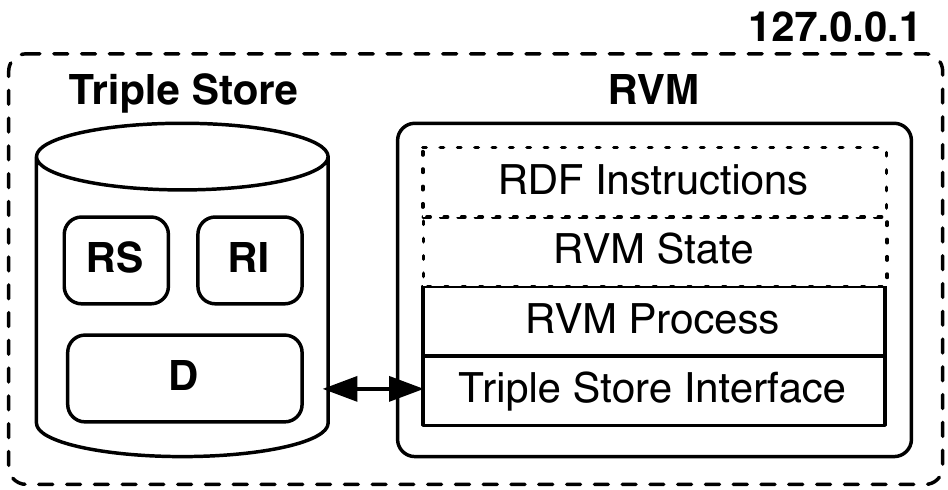}
	\caption{\label{fig:process-layers}The components of an RVM-based computer. The RDF instructions and RVM state are boxed in dotted lines to signify that they can be represented in RDF and thus, able to be placed in a triple store. The double-arrowed line connecting the triple store interface to the triple store is a read/write/delete  protocol such as SPARQL/Update.}
	\end{center}
\end{figure}

An RVM has four primary components: the RDF computing instructions, the RVM state, the RVM process, and a triple store or web server interface.\footnote{The concepts presented in this subsection deal specifically with triple store interfaces only.} All of these components are diagrammed in Figure \ref{fig:process-layers}, where \ttt{RI} represents RDF computing instructions, \ttt{RS} represents the RVM state in RDF, and \ttt{D} represents other, non-procedural triples in the triple store (i.e.~other RDF data).
%%%

The graph of RDF computing instructions that the RVM interprets dictates the evolution of the RVM's state: the instruction being executed, the state of the heap, the state of all the stacks, etc. Because the RVM state can be represented in RDF, it can be distributed in the same manner as any other RDF data (e.g.~as a set of statements in an RDF triple store or as an RDF document on the Web). The role of the RVM process, on the other hand, is to manipulate the RVM state and thereby carry out a computation. RDF data is simply a data structure. Whether that structure includes static or procedural information does not endow it with the ability to compute. In order for that structure to evolve, it relies on some external process to read, write, and delete triples from it. Thus, in order to fully compute, an RVM state relies on an RVM process to alter it. It is through the triple store interface that an RVM process is able to query (i.e.~read) and alter (i.e.~write and delete) an RVM state, computing instructions, and other RDF data. In Figure \ref{fig:process-layers}, the double-arrowed line between the triple store interface and the triple store denotes a read/write/delete protocol such as SPARQL/Update \cite{sparqlupdate:seaborne2007}.

With respect to Neno/Fhat and the other related RDF programming environments (i.e.~Ripple, FABL, Adenosine, and Adenine), \ttt{RI} and \ttt{RS} are located at different levels of abstraction.\footnote{The term ``other" is used to denote programming environments other than Neno/Fhat.} These differences are articulated in the following itemization.
%%%
\begin{itemize}
	\item Other: only \ttt{D} and \ttt{RI} are in the triple store. \ttt{RS} is represented in local memory.
	\item r-Fhat: \ttt{D} is in the triple store, but \ttt{RI} and \ttt{RS} can move between the triple store and the local memory.
	\item Fhat: \ttt{D}, \ttt{RI}, and \ttt{RS} are all contained in the triple store.
\end{itemize}

Theoretically, it is possible to both read RDF computing instructions and change the RVM state while it is represented in the triple store, as in Fhat. However, due to the read/write overhead incurred by such a model, it is preferable to move the instructions and RVM state to local memory for processing. In the Neno/Fhat environment, this is accomplished through the r-Fhat RVM.\footnote{r-Fhat stands for ``reduced Fhat" as it reduces the amount of RDF data being processed by representing the RVM state in the programming constructs of the RVM process.} Also, in Ripple, computing instructions are moved to local memory to increase processing speed.

The remainder of this section will discuss the architecture and instruction set of the Fhat RVM.

\subsubsection{The Fhat RDF Virtual Machine\label{sec:fhat}}

\begin{figure}[h!]
	\begin{center}
		\includegraphics[width=0.5\textwidth]{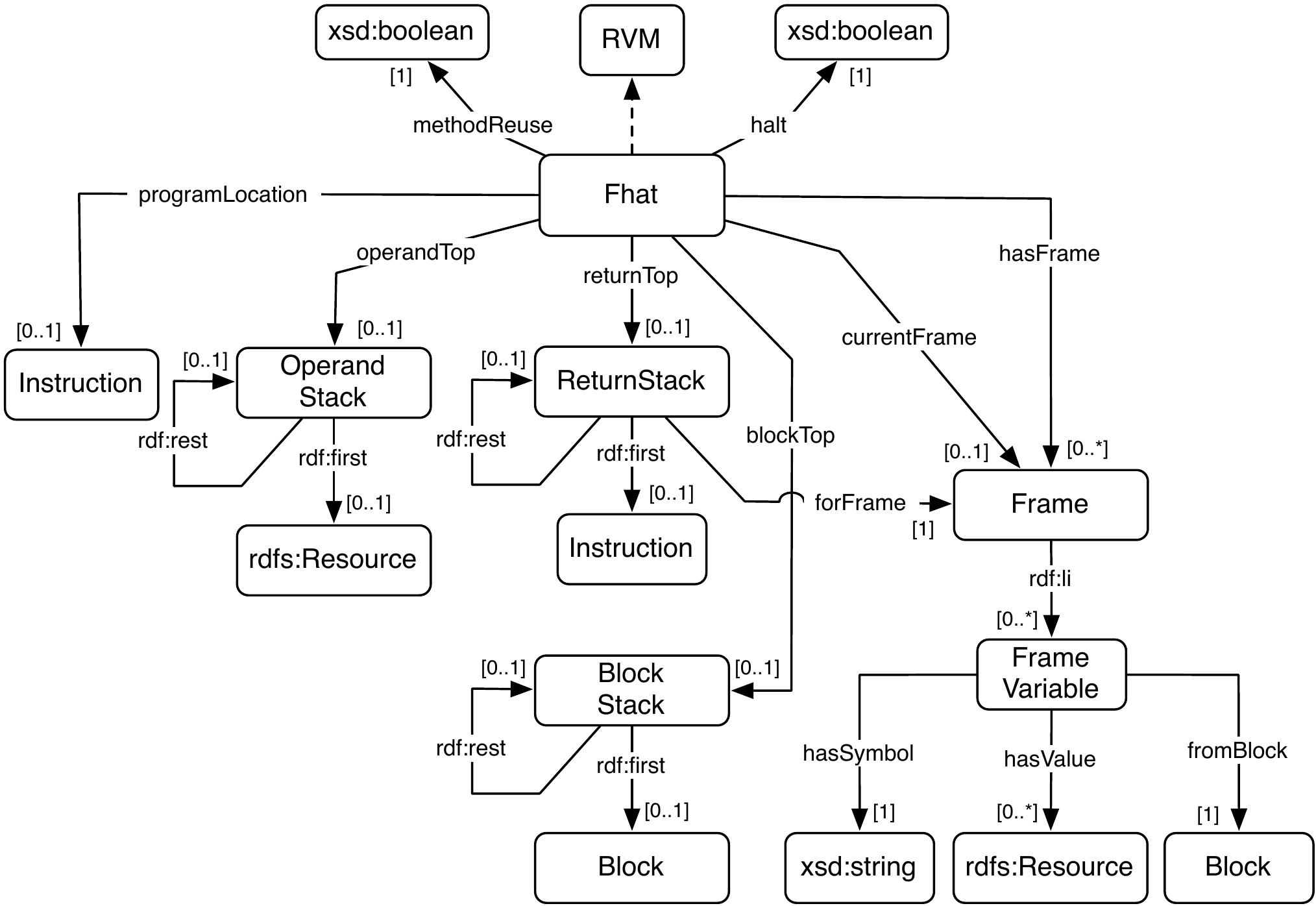}
	\caption{\label{fig:fhat}The classes that compose the Fhat RVM. The dashed lines denote the \ttt{rdfs:subClassOf} property and the bracketed values (e.g.~$[0..1]$) denote the cardinality \ttt{owl:Restriction}s on particular properties. All non-namespaced URIs are part of the \ttt{http://neno.lanl.gov} namespace.}
	\end{center}
\end{figure}

The architecture of the Fhat virtual machine is defined in OWL.\footnote{The Fhat RVM architecture and instruction set are currently available at \url{http://markorodriguez.com/docs/nenoDoc/}.} This architecture provides an abstract description of an instance of a Fhat RVM. Figure \ref{fig:fhat} diagrams the types of resources and relationships present in the Fhat RVM architecture.

\begin{figure}[h!]
	\begin{center}
		\includegraphics[width=0.45\textwidth]{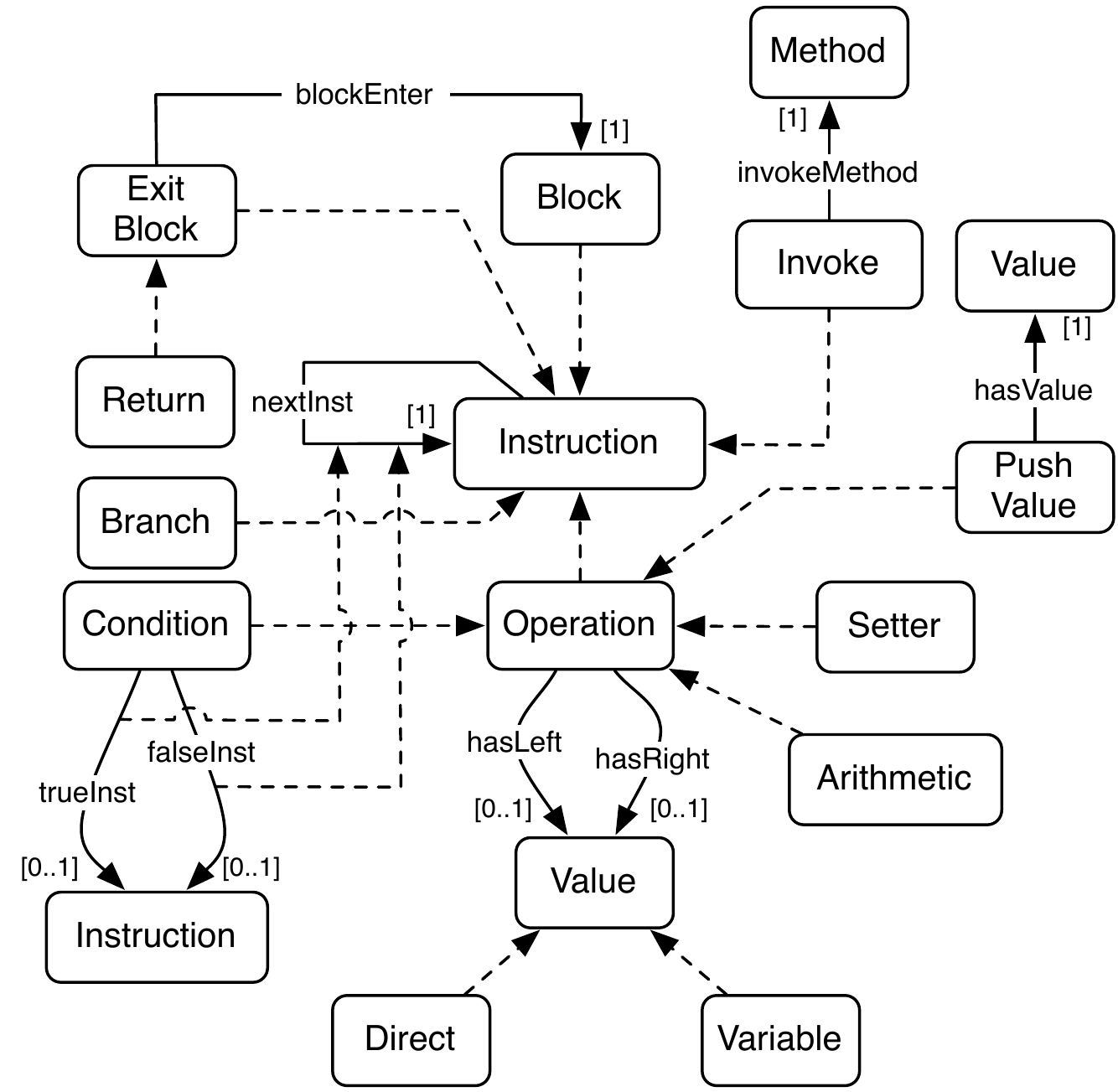}
	\caption{\label{fig:instruction-set}The instruction set of the Fhat RVM. The dashed lines denote the \ttt{rdfs:subClassOf} or \ttt{rdfs:subPropertyOf} properties and the bracketed values (e.g.~$[0..1]$) denote the cardinality \ttt{owl:Restriction}s on particular properties. All non-namespaced URIs are part of the \ttt{http://neno.lanl.gov} namespace.}
	\end{center}
\end{figure}

The Fhat RVM was designed to work with a predetermined set of instructions known as the Fhat instruction set. Figure \ref{fig:instruction-set} diagrams some of the more important instructions supported by the Fhat RVM.

The following itemization presents a few of the more common instructions in the Fhat instruction set and their relationship to the Fhat RVM:\footnote{To preserve the readability of this article, the namespace prefix for the courier font URIs is assumed to be \ttt{http://neno.lanl.gov}.}
%%%
\begin{itemize}
	\item \ttt{Instruction}: A Fhat RVM instance has a \ttt{programLocation} pointer (i.e.~PC) to the current \ttt{Instruction} that it is processing. If no such \ttt{programLocation} exists, then the Fhat RVM can not compute.
	\item \ttt{PushValue}: used to push resources on to the \ttt{OperandStack}. For example, to push the floating point value $2.65$ on the stack for a later operation.
	\item \ttt{Arithmetic}: various subclass instructions include \ttt{Add}, \ttt{Subtract}, \ttt{Multiply}, and \ttt{Divide}. These pop two values off the top of the \ttt{OperandStack}, perform the specified operation, and push the computed value back on the \ttt{OperandStack}.
	\item \ttt{Invoke}: initializes a \ttt{Frame} for a \ttt{Method}. The \ttt{Frame} contains the names (\ttt{hasSymbol}), values (\ttt{hasValue}), and scopes (\ttt{fromBlock}) of the local \ttt{Variable}s of a \ttt{Method}.
	\item \ttt{Setter}: used to assign a value to a \ttt{Variable} in the \ttt{Frame} of a \ttt{Method}. \ttt{SetClear}, \ttt{SetMinus}, \ttt{Set}, and \ttt{SetPlus} are subclasses of \ttt{Setter}.
	\item \ttt{Return}: pushes the return value on the \ttt{OperandStack} and sets the \ttt{programLocation} to the \ttt{Instruction} popped off the \ttt{ReturnStack}. This instruction is used to return from a method.
\end{itemize}

\begin{figure}[h!]
	\begin{center}
		\includegraphics[width=0.485\textwidth]{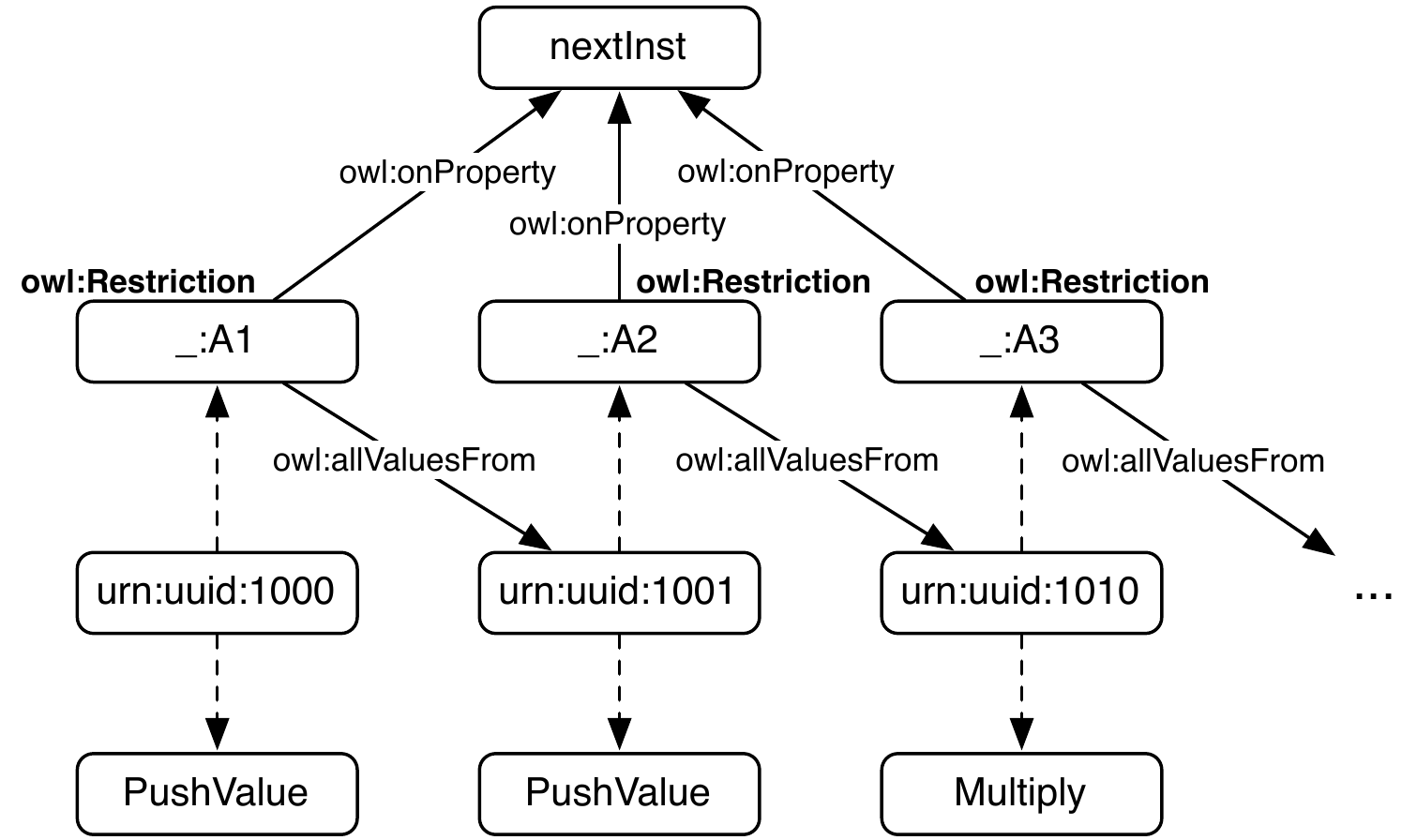}
	\caption{\label{fig:method-api}An snippet of a Fhat API instruction sequence. The dashed lines represent the \ttt{rdfs:subClassOf} property. Note that other \ttt{owl:Restriction}s beside \ttt{\_:A1}, \ttt{\_:A2}, and \ttt{\_:A3} are not presented. For example, a \ttt{PushValue} instruction requires a \ttt{value} to push onto the operand stack. What is presented demonstrates how the sequence of instructions is fixed using \ttt{owl:Restriction}s. All non-namespaced URIs are part of the \ttt{http://neno.lanl.gov} namespace.}
	\end{center}
\end{figure}

When Neno source code is compiled using the Neno/Fhat compiler, a Fhat OWL API is generated. This API provides an abstract representation of a Neno object (known as a Neno class), its fields, its methods, and its method's instructions. The API representation incorporates many \ttt{owl:Restrictions} to ensure that when a Neno object is instantiated, there is an unambiguous generation of instance-level RDF instructions. For example, Figure \ref{fig:method-api} demonstrates how \ttt{owl:Restriction}s are used to define the relationship between instructions within a method body.\footnote{The Neno/Fhat compiler uses Universally Unique Identifiers (UUID) when minting instruction URIs \cite{uuid:leach2005}. UUIDs are 32-bit identifiers. For diagram clarity, only a few characters of a UUID are presented.}
%%%

\begin{figure}[h!]
	\begin{center}
		\includegraphics[width=0.425\textwidth]{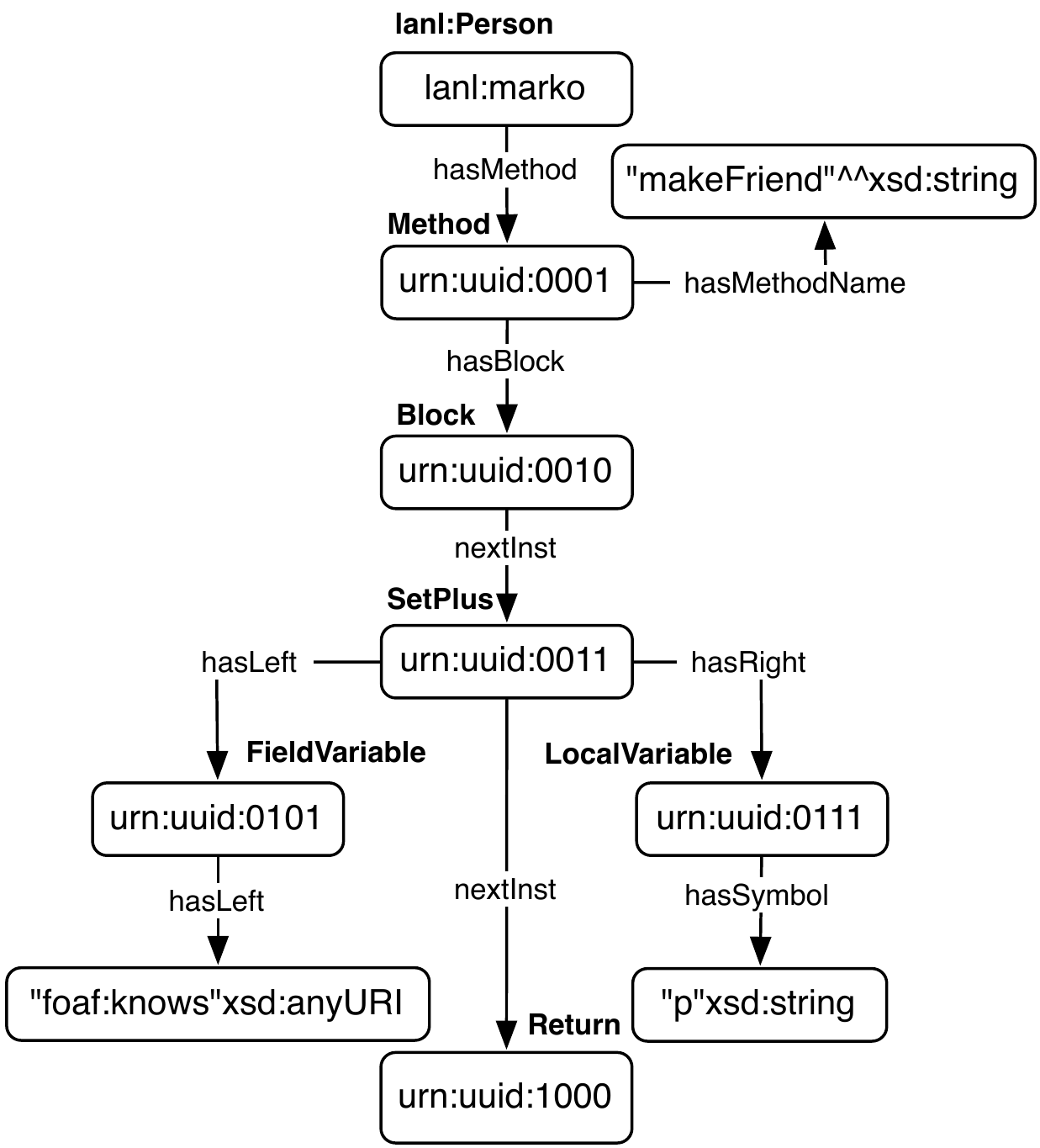}
	\caption{\label{fig:method-example}The instance level RDF representation of the \ttt{makeFriend} method of a \ttt{lanl:Person} class. The bolded terms above the resources denote the \ttt{rdf:type} of the resource. Note that these \ttt{rdf:type}s are inferred types as the direct type is a minted UUID with specific \ttt{owl:Restriction}s as demonstrated in Figure \ref{fig:method-api}. All non-namespaced URIs are part of the \ttt{http://neno.lanl.gov} namespace.}
	\end{center}
\end{figure}

From a Fhat API, it is possible to instantiate Neno objects and their methods. Figure \ref{fig:method-example} diagrams an RDF instance of the \ttt{lanl:Person} \ttt{makeFriend} method previously presented in Section \ref{sec:neno-constructs}. 

\begin{verbatim}
  makeFriend(lanl:Person p) {
    this.foaf:knows =+ p;
  }
\end{verbatim}

In Figure \ref{fig:method-example}, the \ttt{makeFriend} method is associated with a particular \ttt{lanl:Person}, namely \ttt{lanl:marko}. Methods, like RDF properties, are not dependent upon the classes that utilize them in their description. Thus, with Neno it is possible for many objects to share the same method description, or given the requirements of the computation being executed, it is possible for each object instance to have a unique method instance. The latter is desirable when migrating objects between different triple store environments.

\begin{figure}[h!]
	\begin{center}
		\includegraphics[width=0.4\textwidth]{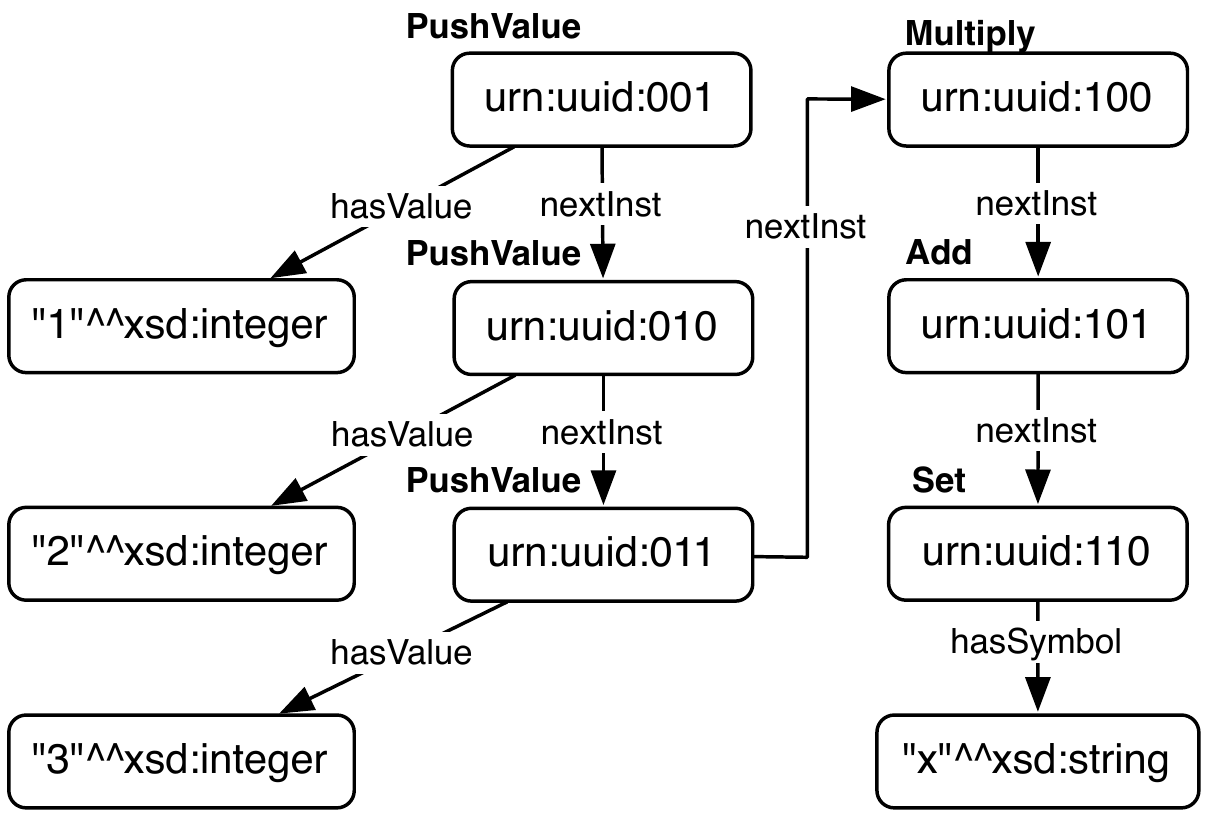}
	\caption{\label{fig:stack-example}An instance of a set of instructions that will set the variable $x$ to the value $1 + (2 \times 3)$. The bolded terms above the resources denote the \ttt{rdf:type} of the resource. All non-namespaced URIs are part of the \ttt{http://neno.lanl.gov} namespace.}
	\end{center}
\end{figure}

Finally, to provide another example of an RDF instruction sequence, Figure \ref{fig:stack-example} diagrams a simple arithmetic instruction sequence that computes $x = 1 + (2 \times 3)$.
%%%

\section{Models of Computing on the Semantic Web}
\label{sec:modelscomputing}

The RVM architecture described in the previous section opens up a number of common computing models to the Semantic Web. The following three models will be discussed throughout the remainder of this section.
%%%
\begin{itemize}
	\item \textbf{Open computing}: an extension of Open Data in which algorithms, virtualized computing machines, and underlying hardware computing resources are made publicly available.\footnote{In another context, the term Open Computing refers to services which allow people to freely use computers in a lab setting. This is not the definition that is used here.}
	\item \textbf{Distributed computing}: a means by which processes are moved to the data, as opposed to the data to the processes.
	\item \textbf{Reflective computing}: as computational process descriptions reside in the URI address space, reflection from the API to the RVM is possible.
\end{itemize}

\subsection{Open Computing}\label{sec:open-computing}

The Open Data movement has ``a philosophy and practice requiring that certain data are freely available to everyone, without restrictions from copyright, patents or other mechanisms of control. Its ethos is similar to that of other open movements and communities such as Open Source and Open Access."\footnote{Quoted from the Wikipedia article on Open Data at \ttt{http://en.wikipedia.org/wiki/Open\_Data}.}

By Open Computing, we understand that
%%%
\begin{itemize}
	\item RDF computing instructions should be made freely available and easily accessible for code reuse, and
	\item results of popular computations should be made publicly available.
\end{itemize}

\subsubsection{Towards a Web of Programs}

A key advantage of the RDF data model is that in using URIs to denote resources, RDF makes resource descriptions distributable in a way which leverages the existing infrastructure of the Web.  For example, suppose that a Semantic Web application encounters the following URI:
%%%
\begin{scriptsize}
\begin{verbatim}
http://www4.wiwiss.fu-berlin.de/factbook/resource/Nepal .
\end{verbatim}
\end{scriptsize}
%%%
If the application is capable of dereferencing the URI over HTTP or querying on it through a SPARQL endpoint, it will find RDF statements about Nepal, including demographic and geographical data.  The application may then use these statements to solve problems. The practice of serving an RDF representation of a resource against its URI, as well as  establishing links between such URIs (for example, \ttt{owl:sameAs} links) is known as Linked Data. Thus, Linked Data is the RDF equivalent of the interlinked hypertext documents which make up the bulk of today's Web. Furthermore, it is the mechanism by which physically isolated RDF data sets are amalgamated into a truly global Web of data.

The distributed nature of Linked Data provides a strong argument for the representation of programs and program state in RDF.  Embedding data structures and algorithms in the web of Linked Data not only makes them universally available, but also eliminates the need for special-purpose software to retrieve and combine programs which reference each other across the underlying physical network.  Effectively, generic Linked Data interfaces such as the Semantic Web Client Library\footnote{Semantic Web Client Library is currently available at \url{http://sites.wiwiss.fu-berlin.de/suhl/bizer/ng4j/semwebclient/}} serve as language-agnostic program linkers, aggregating procedural as well as purely descriptive RDF data dynamically, as needed in a computation. The development of RDF programming languages were motivated by this idea of a ``web of programs". In RDF programming languages, programs become ``first-class" entities in the Linked Data community's distributed graph data structure.  Incorporating them into further programs is as simple as referencing their URIs. 

Similarly, as an alternative to what might be called ``simply" linked programs, encapsulating a computational object in a named graph (as demonstrated later in Section \ref{sec:role-named-graphs} and Figure \ref{fig:named-object}) makes those objects available to any application with knowledge of and access to the graph.  To reference a computational object, an application must be able to infer from the object the URI of the named graph which describes it, as well as the location of a SPARQL endpoint from which it can retrieve the named graph. Emerging solutions such as the Semantic Web Crawling Sitemap Extension\footnote{The Semantic Web Crawling Sitemap Extension is currently available at \url{http://sw.deri.org/2007/07/sitemapextension/}} aid in making this information discoverable by Semantic Web applications.

Finally, in publishing an RDF program to the Semantic Web, it is good practice to provide documentation of the API.  Furthermore, it is natural to express Semantic Web API descriptions in RDF, for instance as OWL ontologies.  Like the JavaDoc framework, OWLDoc\footnote{OWLDoc is currently available at \url{http://www.co-ode.org/downloads/owldoc/}} is a useful aid to developers in learning an OWL API and using it in their applications. The combination of machine-accessible program code and API documentation is particularly appropriate for application scenarios involving the automated discovery and execution of programs.

\subsubsection{Memoization and Computational Reuse}

In some situations, it is best to query for the result of a previous computation than to re-compute it. This idea is known as \textit{memoization} \cite{memo:michie1968}, and the Semantic Web and its open data philosophy provides an ideal medium for such computational reuse. Consider the simple function $f : \mathbb{N} \rar \mathbb{N}$, where $f(n) = n+1$. If $f(5)$ has been previously computed, the result can be represented by the RDF triple $\la 5, \ttt{f}, 6 \ra$.\footnote{For the purpose of this simple example, the caveat that RDF does not allow a literal to be the subject of a triple is ignored.} The results of that computation can be reused by another RVM at a later time. Memoization sacrifices space for time. Of course, this is an impractical example, because re-computing $f(5)$ is faster than querying the Semantic Web for the mapping. However, for other, more computationally complex operations, querying for a result may be orders of magnitude faster than recomputing it. For example, many graph analysis algorithms have a relatively high complexity, such as PageRank (or eigenvector centrality)---$\mathcal{O}(EI)$, closeness centrality---$\mathcal{O}(N^2)$, and betweenness centrality---$\mathcal{O}(NE)$\footnote{This complexity is for unweighted betweenness centrality. Weighted betweenness centrality has a complexity of $\mathcal{O}(NE + N^2 \;\text{log}\; N)$. See \cite{betweeness:brandes2001} for more information on betweenness centrality.}, where $N$ is the number of vertices in the graph, $E$ is the number of edges in the graph, and $I$ is the number of iterations. Such algorithmic complexity become important when considering the use of graph analysis algorithms on the Semantic Web \cite{multigraph:rodriguez2007,grammar:rodriguez2007}. The Semantic Web provides a unique medium by which computations such as these can be stored and later leveraged by other applications. In this sense, not only is metadata open, but so are computational results.

\subsection{Distributed Computing}\label{sec:distributed-computing}

Virtual machine computing provides a layer of abstraction between program instructions and the underlying hardware CPU. It is the role of the virtual machine to serve as a proxy to translate high-level instructions into the respective instruction set of the underlying CPU. While this indirection slows the computation down by incurring a translation step, it permits the same high-level instructions to execute on various physical hardware architectures. This idea is captured in the popular Java slogan of ``write once, run anywhere." In the case of RVMs, this interoperability hides the underlying hardware infrastructure supporting the Semantic Web.

As RDF-based data sets become larger and more numerous, the Semantic Web will be composed of more large-scale RDF repositories serving Linked Data to third-party applications. However, some applications may draw upon more data than can reasonably be moved from server to client. In such instances, it may be worthwhile to migrate an executing program, in the form of RVM state and RDF instructions, to the provider's environment for local processing. The notion of process migration has been proposed to remedy issues surrounding massive-scale data processing \cite{hsieh:migration1993,dean:mapreduce2004} and forms one of the primary purposes of Grid computing \cite{grid:foster2004}.

The remainder of this section illustrates one possible mechanism for the migration of an RVM across different hardware hosts in order to accomplish a distributed computation within the Semantic Web infrastructure. However, before doing so, a discussion of the role of named graphs in distributed Semantic Web computing is required.

\subsubsection{The Role Named Graphs}\label{sec:role-named-graphs}

In a Semantic Web computing environment where instructions, virtual machines, and data commingle within a single RDF data structure, there is an increased need for trust, security, and provenance mechanisms. For example, it may be necessary to 
%%%
\begin{itemize}
	\item group RVM states and RDF instructions for ease of migration and identification,
	\item ``sandbox" RVMs and RDF instructions to prevent malicious or poorly written code from destroying a triple store's data integrity, and
	\item control the permissions that foreign RVMs and RDF instructions have in a triple store environment.
\end{itemize}

The most fundamental construct supporting the above three requirements is RDF reification. Reification provides a way to make statements about statements. In RDF, a reified triple may be the subject or object of another triple. While the concept of RDF reification was initially introduced in the RDF specification with the \ttt{rdf:Statement} construct, recent developments in named graphs (or quads) provide a more manageable solution to triple reification \cite{named:carroll2005}.

In a named graph, a ``triple" is an ordered set of four elements $\la s, p, o, g \ra$.\footnote{While such triples are actually a quads, the term ``triple" will be used as this is a popular convention.}  The $g$ URI, which represents a named graph,  can be used as the subject or object of another statement and thus serves as a mechanism for attaching metadata to the graph. This metadata may include usage statistics, human-level descriptions (e.g.~\ttt{rdfs:comment}), access control permissions \cite{policy:reddivari2005}, and/or provenance information.

\begin{figure}[h!]
	\begin{center}
		\includegraphics[width=0.485\textwidth]{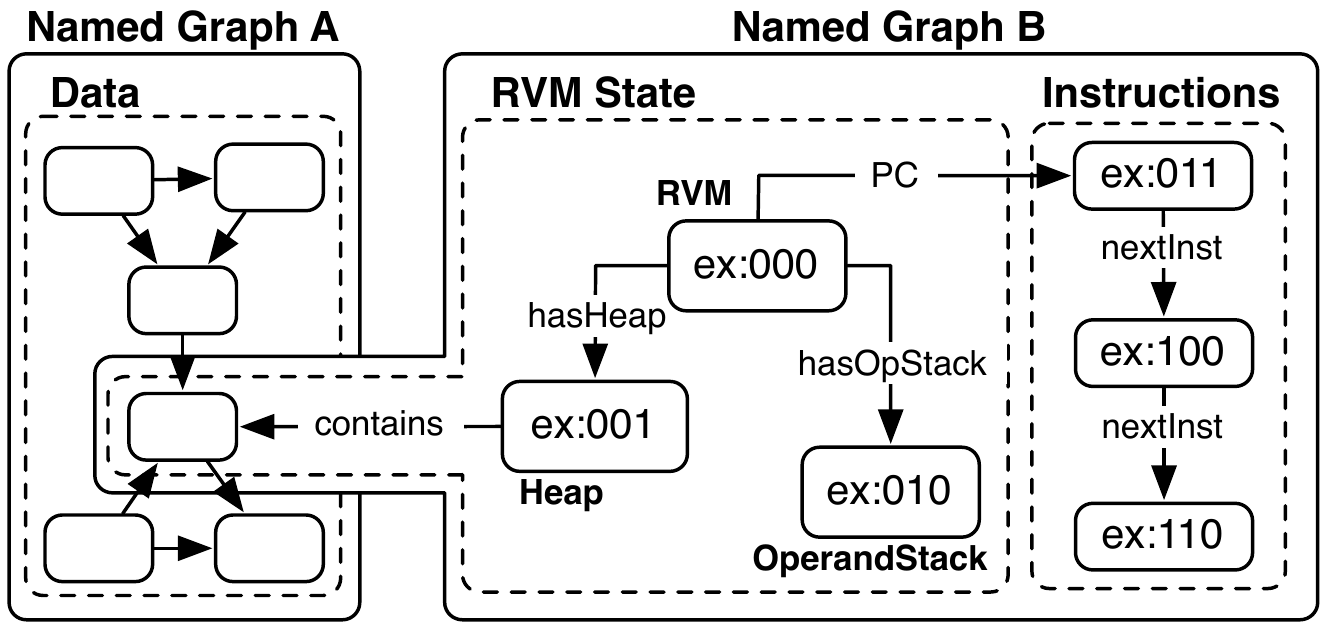}
	\caption{\label{fig:low-level}Using named graphs to encapsulate RVM state and RDF computing instructions. The \ttt{rdf:type} of the RVM state resources is provided in bold next to the resource. All non-namespaced URIs are part of an example RVM namespace denoted \ttt{http://example.com/rvm}.}
	\end{center}
\end{figure}

Figure \ref{fig:low-level} demonstrates how a named graph can encapsulate data, state, and instructions. 
%%%
%%%
It is possible to attach security metadata to named graph \ttt{B} such that the RVM and RDF computing instructions contained in it have specific permissions with respect to the triples in named graph \ttt{A}. It is also straightforward to extract the RVM state and its current instructions by simply selecting triples from named graph \ttt{B}. For example, the query
%%%
\begin{verbatim}
SELECT ?x ?y ?z
  WHERE {
     GRAPH <B> {
    ?x ?y ?z }}
\end{verbatim}
%%%
yields the entire \ttt{B} graph and thus, both the RDF computing instructions and the RVM state.

\begin{figure}[h!]
	\begin{center}
		\includegraphics[width=0.485\textwidth]{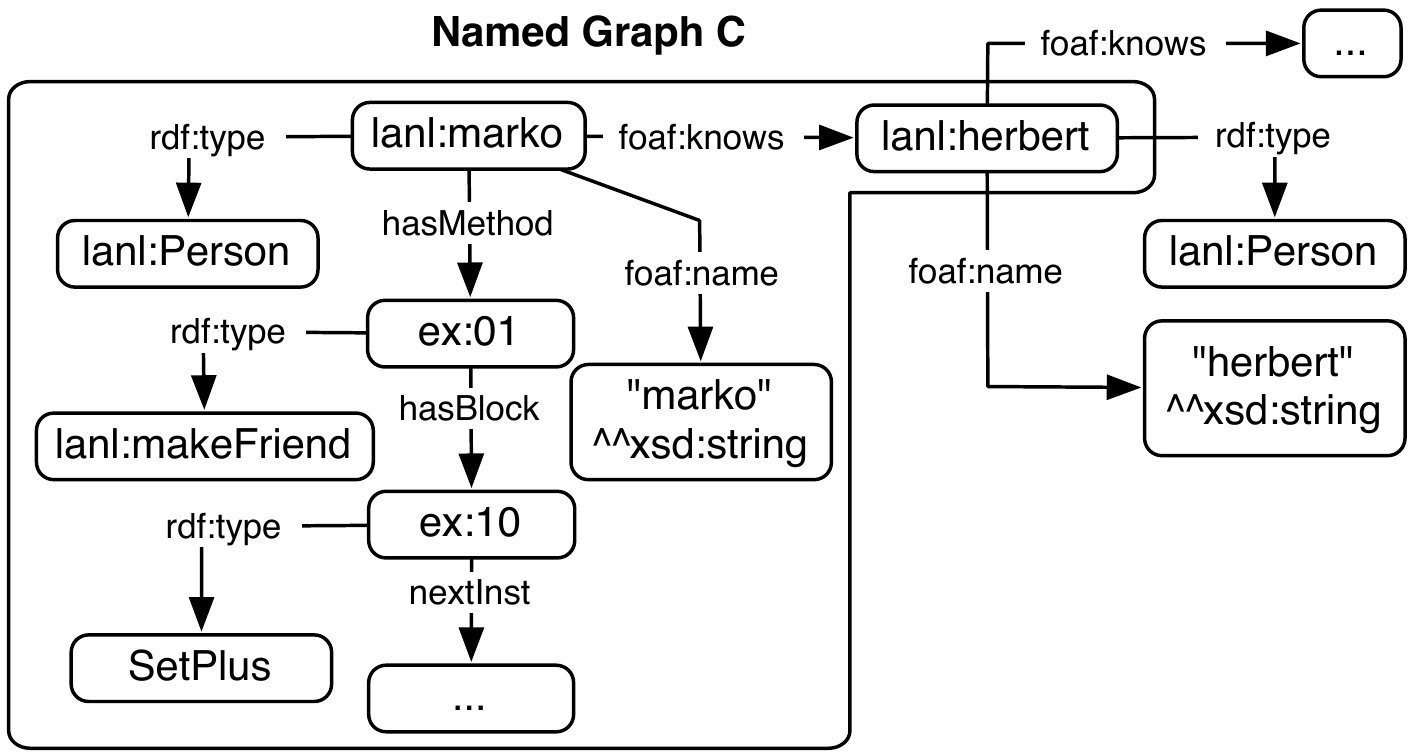}
	\caption{\label{fig:named-object}Using named graphs to encapsulate computational objects. All non-namespaced URIs are part of an example RVM namespace denoted \ttt{http://example.com/rvm}.}
	\end{center}
\end{figure}

The Neno/Fhat programming environment uses named graphs to encapsulate computational objects for ease of migration, code sharing, and internally for garbage collection. For example, in Figure \ref{fig:named-object}, \ttt{lanl:marko} is a URI; however, at a higher-level of abstraction, \ttt{lanl:marko} is a graph-based object with method declarations and explicit method instructions.
%%%

Section \ref{sec:rvm-farm} will now discuss distributed computing on the Semantic Web using named graphs.

\subsubsection{RVM Compute Farms}\label{sec:rvm-farm}

An open hardware provider may use an \textit{RVM farm} to manage concurrent RVMs computing on the local triple store. For example, a Linked Data repository may provide an RVM farm that is used to execute RVMs and instructions that are working with the data currently in its repository. An RVM farm polls its associated triple store for non-executing RVM states. Once a non-executing state has been found, the RVM farm will spawn an RVM process to execute it. For instance, an RVM farm might use the query
%%%
\begin{footnotesize}
\begin{verbatim}
SELECT ?x 
  WHERE { 
    ?x <rdf:type> <rvm:RVM> .
    ?x <rvm:needsProcess> "true"^^xsd:boolean }
\end{verbatim}
\end{footnotesize}
%%%
to locate RVM states that need processing. If a URI binds to \ttt{?x}, an RVM process is created. The newly created RVM process is passed the bound \ttt{?x} URI as a parameter. Thus, the newly created RVM process knows which RVM state to harvest and execute. An RVM process maintains no state information and thus, if an RVM process halts (for whatever reason), the current state of computation is maintained in the RVM state. That is, the state of the stacks, program counter, etc. are frozen until another RVM process can continue its execution. In this way, RVM state encoding makes it desirable for migration between triple store environments, and thus, RVM farms.

\begin{figure}[h!]
	\begin{center}
		\includegraphics[width=0.485\textwidth]{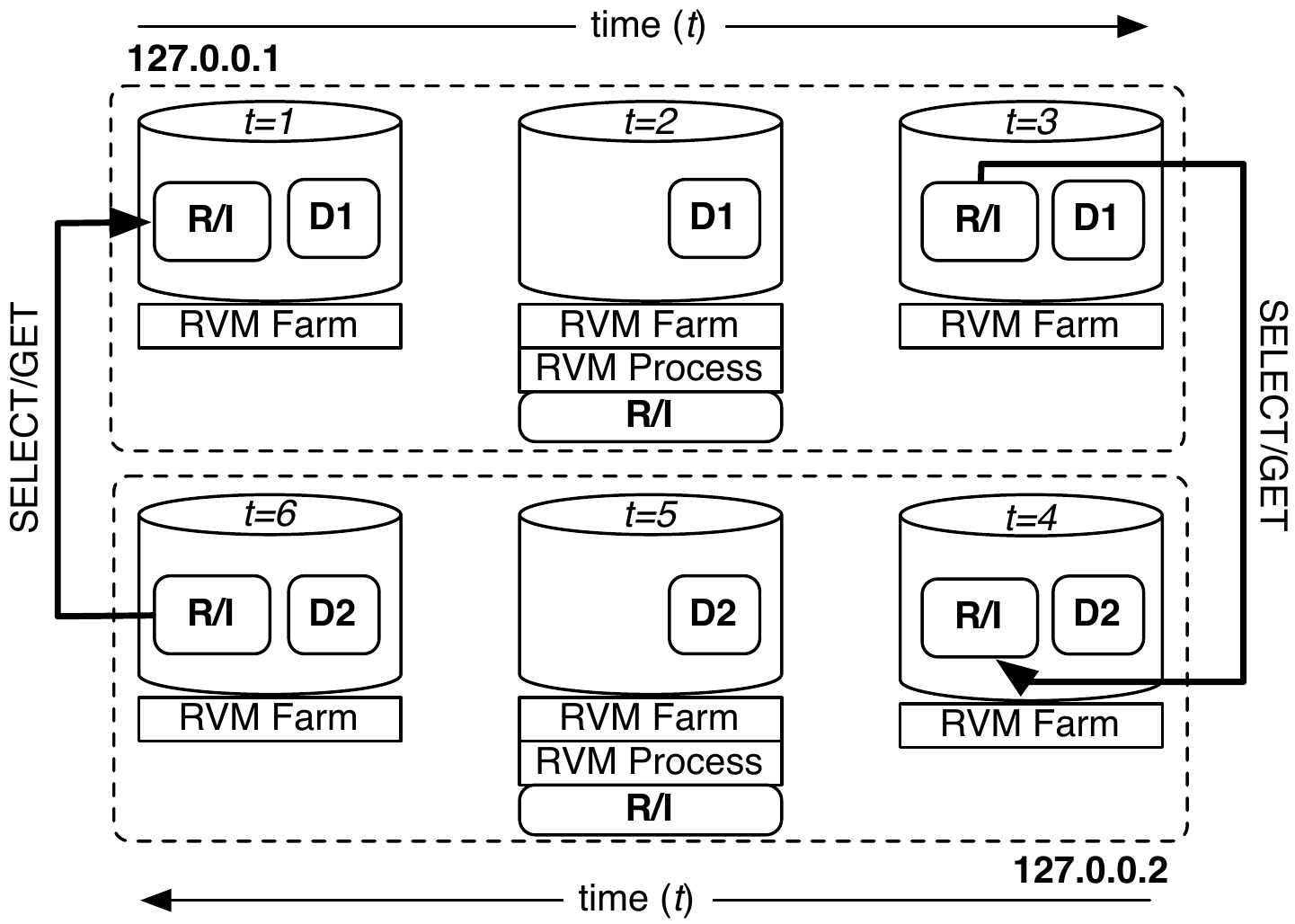}
	\caption{\label{fig:migrate}Migrating RVM and RDF computing instructions between host triple stores.}
	\end{center}
\end{figure}

Figure \ref{fig:migrate} diagrams a migration pattern between two triple store environments, where one environment is located at \ttt{127.0.0.1} and the other is located at \ttt{127.0.0.2}.

Starting at $t=1$ in Figure \ref{fig:migrate} (top left corner), an RVM state and RDF instructions named graph (denoted \ttt{R/I}) exist in the \ttt{127.0.0.1} triple store. At $t=2$, the local RVM farm locates the RVM state and spawns a new RVM process. The RVM process moves the RVM state and instructions to local memory to perform a computation.\footnote{Moving the RVM state and instructions to memory ensure a more efficient use of clock cycles. This was articulated previously when discussing the r-Fhat RVM.} Assume that the RVM uses the data in the named graph denoted \ttt{D1} in its computation. At $t=3$ the RVM has finished its computation with \ttt{D1} and inserts its state and instructions back into the \ttt{127.0.0.1} triple store. The RVM process at \ttt{127.0.0.1} notifies the RVM farm at \ttt{127.0.0.2} that it has an RVM that wishes to migrate to its triple store. The RVM farm at \ttt{127.0.0.2} then harvests the RVM and its intructions using a SPARQL SELECT query or HTTP GET request.\footnote{Another approach would be to have the RVM process at \ttt{127.0.0.1} INSERT/PUT the RVM state and instructions into the triple store at \ttt{127.0.0.2}.} Thus, what is migrated is the RDF instructions as well as the state of the RVM which includes, amongst other information, a populated heap that it is using in its computation.

At $t=4$, the \ttt{R/I} named graph is located in the triple store at \ttt{127.0.0.2}. At $t=5$, the RVM farm at \ttt{127.0.0.2} locates \ttt{R/I} and then spawns an RVM process. The newly created RVM process moves the RVM state and instructions to main memory for local processing on the \ttt{D2} named graph data set. When the RVM no longer requires \ttt{D2} for its computation, it can be moved back to the \ttt{127.0.0.2} triple store at $t=6$ and ultimately, migrate to yet another triple store at $t=7$.

Such a model of distributed computing is dependent on mechanisms of trust and security on the Semantic Web. A simple solution would be to allow a foreign RVM to read, write, or delete from its own named graph and any named graphs that it spawns, but to allow it only to read from other named graphs in the triple store. Furthermore, limiting the number of triples in a named graph can prevent the creation of an excessive amount of data by a foreign RVM. Finally, in Neno/Fhat, a simple ``halt" mechanism regulates the number of clock cycles that an RVM state can utilize.

\subsection{Reflective Computing}\label{sec:reflective-computing}

The concept of reflection in computing refers to the ability for software to modify itself during runtime \cite{maes:reflection1987,demers:reflection1995}. For instance, in the Java programming language, it is possible for an object to ``look at" the API at runtime and make choices as to flow of execution. This is made possible through the \ttt{java.lang.reflect} package of the core Java API. This type of reflection exists because the description of an object is at the same level of abstraction as the object itself. In general, reflective computing is possible when particular states of a program are made available within a representation that is processable by the executing program. In many respects, reflective computing draws many parallels to the common model of reification in logic, and specifically in RDF using named graphs or the \ttt{rdf:Statement} construct. With reification, it is possible to make a statement about a statement. With reflection, it is possible to compute with the description of a computation -- whether that description is of a particular state of computation or of the program itself \cite{reflection:sobel1996}. 

Reflective computing has found application in heterogeneous object environments where an object may need to discover new objects and ``learn" what functionality they have or reason about their functionality before leveraging them within a computation. OWL-S \cite{owls:martin2004} and other web service description frameworks serve a similar purpose in that they provide detailed machine-readable descriptions of the input requirements, processing stages, and ultimate output of a service. With respect to RDF-based programming languages and RDF-encoded RVMs, both procedural and machine information are encoded at the same level of abstraction, namely in RDF, thus making them readily available for run-time analysis. While programming constructs such as packages, class descriptions, methods, and so forth are utilized for the purpose of procedural encapsulation, it is possible to make use of queries on and manipulations of an RDF graph in order to reason on and alter the full computing stack during run-time. This sub-section will present three types of reflection that are made salient in RDF computing: object and method reflection, instruction reflection, and machine reflection. 

\subsubsection{Object and Method Reflection}

A triple store supports three basic operations: read, write, and delete. RDF programming languages make use of these operations to query and manipulate an RDF graph in order to evolve the RDF graph and thus, compute. It is possible for an RDF program to query a triple store in order to retrieve information about the state of an object whether that object be itself or another object. For example, an object might execute a query to locate all objects of \ttt{rdf:type} \ttt{lanl:Person} as well as any methods associated with that object. The following SPARQL query returns a list of instantiated \ttt{lanl:Person} objects and their associated method URIs:
%%%
\begin{verbatim}
SELECT ?x ?y 
  WHERE {
    ?x <rdf:type> <lanl:Person> .
    ?x <rvm:hasMethod> ?y }.
\end{verbatim}
%%%
Given the URIs bound to \ttt{?x} (the URI \ttt{lanl:Person} resources) and \ttt{?y} (the URI of the \ttt{lanl:Person} methods), the querying object can make decisions as to how to utilize these URIs in its processing. For example, it could decide that if a particular \ttt{lanl:Person} resource has a \ttt{makeFriend} method, then it must be a ``friendly" object and will make friends with that object as well as invoke that object to make friends with it. Thus, creating a symmetric \ttt{foaf:knows} relationship. 

This type of reflection is analogous to class reflection in Java. For example, in Java, the previous method reflection query may be executing as
%%%
\begin{verbatim}
Method[] methods = Person.getMethods();
\end{verbatim}
%%%
The class \ttt{Person} is queried for its set of methods which are returned as an array of \ttt{Method} objects. These \ttt{Method} objects can then be computed with like any other object. For example,
%%%
\begin{verbatim}
Person marko = new Person();
Method[] methods =
    josh.getClass().getMethods();
for(Method m : methods) {
  if(m.getName().equals("makeFriend")) {
    marko.makeFriend(josh);
    m.invoke(josh, marko);  
  }
}
\end{verbatim}
%%%
Both \ttt{makeFriend} and \ttt{invoke} are presented in the last two instruction lines to demonstrate the two ways in which the same method can be executed.

\subsubsection{Instruction Reflection}

In an RDF computing environment, not only are methods exposed, but so are the instructions that composes those methods. What is returned by \ttt{Person.getMethods()} in the previous example is an array of pointers to the methods that are available from that class. In this way, the program, at run-time, is able to inspect the \ttt{Person} Java API. Once this method pointer has been acquired, in Java, it is possible to invoke the method:
%%%
\begin{verbatim}
Person marko = new Person();
methods[0].invoke(marko, null);
\end{verbatim}
%%%
In an RDF computing environment this is equivalent to adding an \ttt{Invoke} instruction resource as the next instruction in the current instruction sequence. This ensures that the next instruction to be processed by the executing RVM will invoke the method. Assuming \ttt{y} is the URI of the \ttt{?y} binding of the previous SPARQL select query, the following SPARQL/Update command will provide the appropriate alteration of the flow of execution:
%%%
\begin{verbatim}
INSERT DATA {
  <ex:001> <rvm:nextInst> <ex:010> .
  <ex:010> <rdf:type> <rvm:Invoke> .
  <ex:010> <rvm:invokeMethod> <y> }.
\end{verbatim}

As demonstrated, it is possible to reason about the current instruction sequence of a program and perhaps, insert new instructions as a result. This type of direct code manipulation supports evolutionary (or genetic) computing: at runtime, new code can be introduced into the system. Again, the runtime creation and manipulation of code is made possible by the fact that the API and the instructions are at the same level of abstraction: URIs, literals, blank nodes, and triples.

\subsubsection{Machine Reflection}

\begin{figure}[h!]
	\begin{center}
		\includegraphics[width=0.485\textwidth]{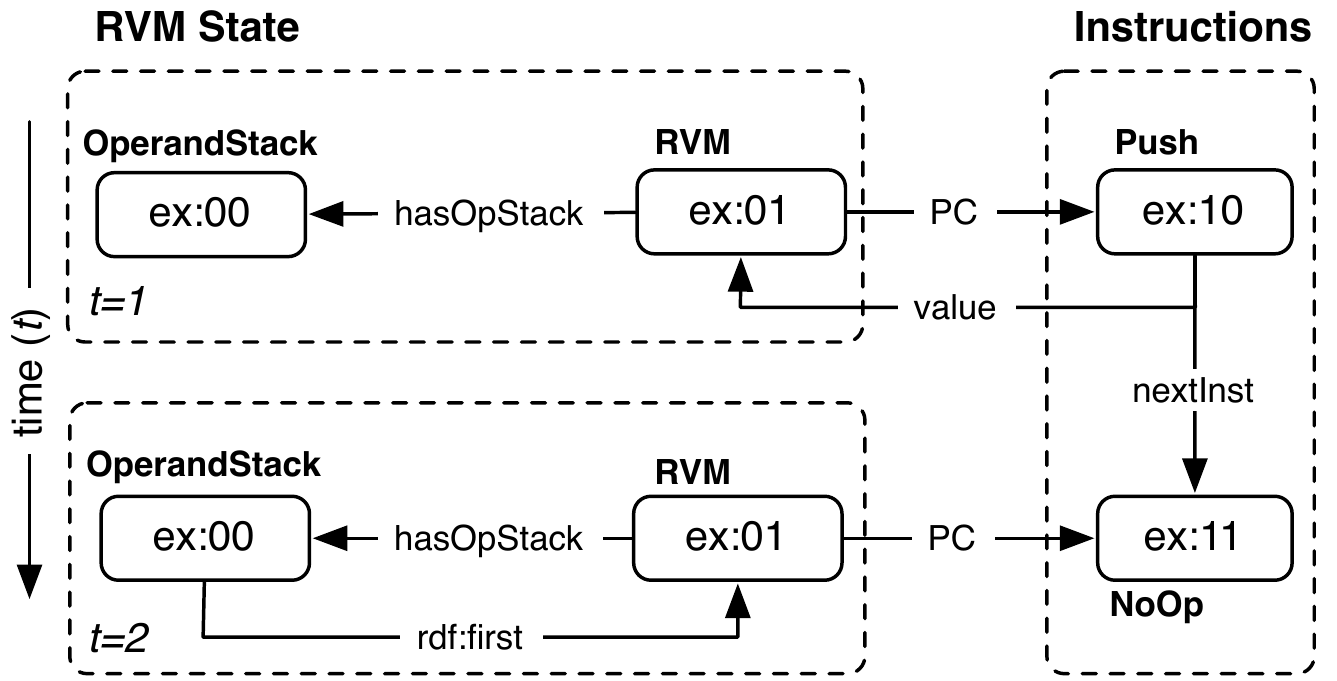}
	\caption{\label{fig:reflection}RDF reflection at the RVM-level. Time is only specified for the RVM state, not the instructions. The \ttt{rdf:type} of the resource is presented in bold with the resource. All non-namespaced URIs are part of an example RVM namespace denoted \ttt{http://example.com/rvm}.}
	\end{center}
\end{figure}

An RVM may execute instructions that manipulate itself. Thus, a machine can modify itself at runtime. This type of machine reflection is diagrammed in Figure \ref{fig:reflection}.

In Figure \ref{fig:reflection}, at $t=1$, the RVM (\ttt{ex:01}) is pointing to an instruction (represented by the program counter \ttt{PC} property) that requires the RVM to push a pointer of itself onto its own operand stack (i.e.~the \ttt{Push} instruction's \ttt{value} is the \ttt{RVM} resource). At $t=2$, the RVM URI is on the top of the operand stack (i.e.~\ttt{rdf:first}). Thus, at this point, any instruction that utilizes the operand stack will involve the RVM computing on itself (in Figure \ref{fig:reflection} this next operation in a \ttt{NoOp}, which will not alter the state of the machine). In this way, it is possible for the RVM to manipulate itself at runtime. If the machine's process is also encoded in RDF (e.g.~by coding Fhat's process in Neno), then the complete RVM is subject to such machine-level reflection. In short, the RVM's stacks, frames, program counter, or whichever modeled components the RVM in question represents can be manipulated through machine-level reflection.

The previous example is an extreme case of reflection which is not found in typical programming environments.  For instance, in Java, it is not possible for a program to obtain a pointer to the JVM. Furthermore, the JVM is represented according to the instruction set of the underlying hardware CPU and thus, is not at the same level of abstraction. This reduces the ability to facilitate machine reflection. In languages which compile to machine-specific code, such as C and C++, it is possible to get a direct pointer to the program being executed. In this way, a program may be manipulated at runtime. 

In an RDF-based programming language, reflection can be applied to the entire computational stack. However, it is still possible to engineer code that respects the common principles of data hiding, encapsulation, and modularity. In many cases, such common development practices are preferable. The purpose of this section was to demonstrate the flexibility of this RDF-based programming style.

In a Semantic Web computing environment where object persistence is an expected feature, the ability for objects to discover, reason, and ultimately interact with other objects will be a key component of RDF-based applications. Moreover, with persistent RVMs, it will be important for RVM processes to discover and reason on RVM states before executing them.

\section{Conclusion}

The Semantic Web is a distributed environment in which descriptive world-models can be queried and manipulated by external applications. These external applications leverage the world-wide repository of structured multi-relational data for the purpose of computation. This article has addressed the potential role of encoding such applications in the Semantic Web. Given the modeling power of RDF, it is possible to represent not only data, but also software and virtual machines in RDF. RDF-based programming languages were designed to take explicit advantage of the RDF data model. Unlike RDF APIs in other languages such as Java, C, etc., these languages do not require the developer to work with two different data models. There is no disjoint experience for the developer \cite{fabl:bureau2001}. Furthermore, with the ability to encode virtual machines and their state in RDF, it is possible to migrate software and machines to other data sets around the world. This provides a distribute process infrastructure to the Semantic Web's distributed data structure. When the more procedural aspects of computing are embedded in the Semantic Web, new computing models emerge that push the Semantic Web towards a distributed general-purpose computing environment.

\section{Acknowledgements} 

Joshua Shinavier of Knowledge Reef Systems Inc. and the Rensselaer Polytechnic Institute and the designer of the Ripple RDF programming language provided much insight during the development of these ideas. Moreover, Josh contributed helpful text to an original draft version. Joe Geldart of the University of Durham and designer of the Adenosine RDF programming language and Callaghan RVM provided assistance in the discussion of Adenosine. Likewise, Chris Goad, the designer of the FABL language and RVM provide assistance with FABL. Ryan Chute of the Digital Library Research and Prototyping Team of the Los Alamos National Laboratory engineered the Neno/Fhat compiler and Fhat RVM process. This work was supported by both the Digital Library Research and Prototyping Team and the Center for Nonlinear Studies of the Los Alamos National Laboratory.

%\balancecolumns

\end{document}